\documentclass[fleqn,usenatbib]{mnras}

\usepackage{newtxtext,newtxmath}
\usepackage[utf8]{inputenc}
\usepackage{fix-cm}

\usepackage[T1]{fontenc}

\DeclareRobustCommand{\VAN}[3]{#2}
\let\VANthebibliography\thebibliography
\def\thebibliography{\DeclareRobustCommand{\VAN}[3]{##3}\VANthebibliography}

\usepackage{amsmath}
\usepackage{bm}
\usepackage{geometry} 
\usepackage{longtable}
\usepackage{subcaption}
\usepackage{booktabs}
\usepackage{colortbl} 
\usepackage[dvipsnames]{xcolor} 
\usepackage{pdflscape}
\usepackage{float}
\usepackage{placeins}
\usepackage{comment}
\usepackage{adjustbox}
\usepackage{graphicx}
\usepackage{color}

\newcommand{\tess}{\emph{TESS}}

\title[Orbital period catalogue of CVs from TESS]{A Catalogue of Orbital Periods of Cataclysmic Variables and Candidates from \textit{TESS} Observations}

\author[M. K. Dağ et al.]{Meryem K. Dağ\thanks{E-mail: meryem.k.dag@durham.ac.uk}$^{1,2}$,
Simone Scaringi$^{1,3}$,
Kieran O'Brien$^{2}$,
Martina Veresvarska$^{1}$,
Nikita Rawat$^{4}$,
\newauthor
Yusuke Tampo$^{4,5}$,
Santiago Hernández-Díaz$^{6}$,
Colin Littlefield$^{11,12}$,
Krystian I\l{}kiewicz$^{7}$,
Domitilla de Martino$^{3}$,
\newauthor
D.\,A.\,H. Buckley$^{4,5,9}$,
Zackery A. Irving$^{8}$,
Liliana E. Rivera Sandoval$^{10}$,
Wendy Mendoza$^{10}$,
Ryan J. Oelkers$^{10}$,
\newauthor
Peter Garnavich$^{13}$,
Gavin Ramsay$^{14}$,
Yuri Cavecchi$^{15,21,22}$,
Manuel Pichardo Marcano$^{16}$,
J. Kára$^{10}$,
\newauthor
Elmé Breedt$^{17}$,
Axel D. Schwope$^{20}$,
Christian Knigge$^{18}$,
N.Castro Segura$^{19}$,
Boris G\"ansicke$^{23}$,
Keith Inight$^{23}$
\\
$^{1}$Centre for Extragalactic Astronomy, Department of Physics, Durham University, South Road, Durham, DH1 3LE, UK\\
$^{2}$Centre for Advanced Instrumentation, Department of Physics, Durham University, South Road, Durham, DH1 3LE, UK\\
$^{3}$INAF -- Osservatorio Astronomico di Capodimonte, Salita Moiariello 16, I-80131 Naples, Italy\\
$^{4}$South African Astronomical Observatory, PO Box 9, Observatory 7935, Cape Town, South Africa\\
$^{5}$Department of Astronomy, University of Cape Town, Private Bag X3, Rondebosch 7701, South Africa\\
$^{6}$Institut für Astronomie und Astrophysik, Eberhard Karls Universität Tübingen, Sand 1, 72076 Tübingen, Germany\\
$^{7}$Nicolaus Copernicus Astronomical Centre, Polish Academy of Sciences, Bartycka 18, PL-00-716 Warszawa, Poland\\
$^{8}$School of Physics and Astronomy, University of Southampton, University Road, Southampton SO17 1BJ, UK\\
$^{9}$Department of Physics, University of the Free State, PO Box 339, Bloemfontein 9300, South Africa\\
$^{10}$Department of Physics and Astronomy, University of Texas Rio Grande Valley, Brownsville, TX 78520, USA\\
$^{11}$NASA Ames Research Center, Moffett Field, CA 94035, USA\\
$^{12}$Bay Area Environmental Research Institute, Moffett Field, CA 94035, USA\\
$^{13}$Department of Physics and Astronomy, University of Notre Dame, Notre Dame, IN 46556, USA\\
$^{14}$Armagh Observatory and Planetarium, College Hill, Armagh, BT61 9DG, N. Ireland, UK\\
$^{15}$Departament de Fis\'{i}ca, EEBE, Universitat Polit\`ecnica de Catalunya, Av. Eduard Maristany 16, 08019 Barcelona, Spain\\
$^{16}$Universidad Nacional Autónoma de México, Instituto de Astronomía, Ciudad Universitaria, 04510 Ciudad de México, México\\
$^{17}$Institute of Astronomy, University of Cambridge, Madingley Road, Cambridge CB3 0HA, UK\\
$^{18}$School of Physics and Astronomy, University of Southampton, Highfield, Southampton SO17 1BJ, UK\\
$^{19}$Astronomy and Astrophysics Group, Department of Physics, University of Warwick, Coventry CV4 7AL, UK\\
$^{20}$Leibniz-Institut für Astrophysik Potsdam (AIP), An der Sternwarte~16, 14482 Potsdam, Germany\\
$^{21}$ Departamento de Astrof\'isica, Universidad de La Laguna, 38206, San Crist\'obal de La Laguna, Tenerife, Spain\\
$^{22}$ Instituto de Astrof\'{i}sica de Canarias, 38205 San Crist\'obal de La Laguna, Tenerife, Spain\\
$^{23}$ Department of Physics, University of Warwick, Coventry CV4 7AL, UK}

\date{Accepted XXX. Received YYY; in original form ZZZ}

\pubyear{\the\year{}}

\begin{document}
\label{firstpage}
\pagerange{\pageref{firstpage}--\pageref{lastpage}}
\maketitle

\begin{abstract}

We present a systematic analysis of 2544 cataclysmic variable systems and related candidates observed by the Transiting Exoplanet Survey Satellite (\textit{TESS}), with the aim of compiling a comprehensive catalogue of orbital periods. Using 2-minute photometric time-series data, we applied an automated algorithm to generate Lomb–Scargle periodograms and identify the most significant coherent periodic signals, which were subsequently verified through visual inspection. This process yielded a confident sample of 910 sources exhibiting at least one periodic signal, hereafter referred to as the Cataclysmic Variable Confident Catalogue (CCC). For each object, we report the most likely orbital period together with additional periodic features such as spin modulations and/or superhump signals when present. To assess consistency with previously published measurements, we cross-matched the CCC with the Ritter \& Kolb catalogue, identifying 300 overlapping systems, of which 215 showed full agreement with the R\&K orbital periods, while 39 displayed discrepancies for which the R\&K values were revised based on our TESS measurements and supporting evidence from the literature. Overall, the CCC provides a means to validate known orbital periods, propose corrections where necessary, and offer new determinations for systems with previously unknown periods, thereby supporting a more complete and reliable characterisation of the cataclysmic variable population.


\end{abstract}

\begin{keywords}
cataclysmic variables -- binaries: close -- catalogues
\end{keywords}



\section{Introduction}

Cataclysmic variables (CVs) are interacting binary systems in which a white dwarf accretes matter from a donor star, typically a late-type (K or M) main sequence-like star. As the orbital period of the system decreases to several hours, angular momentum losses drive the donor star to fill its Roche lobe and begin transferring material through the inner Lagrangian point (L1), forming an accretion disk around the white dwarf \citep{Warner1995, Hellier2001}.Although most CVs have orbital periods below a few hours, systems with significantly longer periods also exist. These long-period CVs generally contain more massive or slightly evolved donor stars, which are able to fill their Roche lobes only at larger orbital separations, resulting in higher mass ratios and sustained mass transfer \citep{Podsiadlowski2003}.

CV systems are classified into subtypes based on the magnetic field strength of the white dwarf, the mass transfer rate, and observational characteristics such as the geometry of the accretion flow and variability behaviour \citep{Warner1995}. Considering that all astrophysical systems possess some degree of magnetic field, the group of CVs commonly referred to as non-magnetic generally includes systems in which the white dwarf magnetic field is too weak to significantly affect the
dynamics of the accretion flow. Conversely, magnetic CVs are those in which the field is strong enough to directly influence the accretion process, typically corresponding to surface magnetic field strengths of order $B \gtrsim 10^6$~G, although the exact boundary also depends on the mass-transfer
rate \citep{Warner1995}. In these systems, the magnetic field of the white dwarf disrupts or suppresses the accretion disc, channeling the infalling material along magnetic field lines toward the magnetic poles. Magnetic CVs are traditionally subdivided into polars (AM~Her stars) \citep{Cropper1990} and intermediate polars (DQ~Her stars)
\citep{patterson1994dqherc}, depending on the strength of the magnetic coupling between the white dwarf and the accretion flow. In polars, the magnetic field is sufficiently strong to prevent the formation of an accretion disc entirely,
resulting in direct magnetically controlled accretion. The strong coupling between the magnetic field and the accretion flow leads to synchronisation between the
white dwarf spin and the orbital period, with typical surface field strengths of order $B \sim 10^7$--$10^8$~G \citep{Cropper1990}. In contrast, intermediate polars host weaker magnetic fields and are therefore not fully synchronised. In these systems, a truncated accretion disc may still form, while the inner regions of the accretion flow are magnetically channelled onto the white dwarf’s magnetic poles. As a result, the white dwarf typically spins faster than the binary orbit, giving rise to a coherent spin modulation observable at a frequency distinct from the orbital signal \citep{patterson1994dqherc, Hellier2001}.


Non-magnetic systems, by contrast, are further divided into two main observational categories: dwarf novae (DN) and nova-like (NL) variables \citep{Warner1995, osaki1996}. DN systems exhibit semi-periodic outbursts driven by thermal-viscous instabilities in the accretion disk, whereas NL systems show high and relatively steady mass transfer rates without outbursts \citep{Warner1995, osaki1996}. Physically, this distinction is generally understood in terms of the mass-transfer rate from the donor star: systems with mass-transfer rates below the critical value predicted by the disk instability model (DIM) develop thermally unstable accretion disks and appear as dwarf novae, while systems with transfer rates above this critical threshold maintain thermally stable disks and are observed as nova-like variables.In addition, classical and recurrent novae, resulting from thermonuclear explosions on the surface of the white dwarf, form another distinct subclass of CVs \citep{starrfield2008thermonuclear}. 

A related (but somewhat separate) class of systems is the AM Canum Venaticorum systems (AM CVns). AM CVn systems are ultra-compact binaries consisting of a white-dwarf primary and a helium-rich donor star. They exhibit extremely short orbital periods of approximately 5–65 minutes \citep{Solheim, 2020MNRAS.496.1243G, 2025A&A...700A.107G}. The degenerate structure of the donor implies a very small stellar radius, allowing the binary to sustain stable mass transfer within an extremely compact Roche lobe and naturally explaining the very short orbital periods observed in these systems, which are well below the minimum period of hydrogen-rich cataclysmic variables. Helium transferred from the donor forms an accretion disk around the white dwarf, and the absence of hydrogen lines, together with the dominance of helium in their spectra,
constitutes one of the most distinctive observational characteristics of these systems \citep{Ramsay2018_Gaia}.

An important observational diagnostic used in classification is the
detection of superhumps, which are photometric modulations with
periods slightly longer or shorter than the orbital period
\citep{Warner1995, Hellier2001, kato2009superhumps, Bruch2023, Sun2024}. Positive superhumps, with periods
slightly longer than the orbital period, arise from the apsidal
precession of an eccentric accretion disk that has undergone a
3:1 orbital resonance \citep{whitehurst1988, osaki1996,
voloshina2016superhumps, Bruch2023}. The observed superhump period does not correspond to the
disk precession period itself, but instead represents the beat
frequency between the orbital motion and the slow apsidal precession
of the disk, which typically occurs on timescales of several days. In contrast, the origin
of negative superhumps, with periods slightly shorter than the
orbital period, is less well determined. They are generally thought
to result from the retrograde precession of a tilted accretion disc
\citep{patterson1999superhumps, wood2000superhumps, Sun2024}. As in the positive case, the observed
photometric modulation corresponds to the beat between the orbital
period and the nodal precession period, the latter typically ranging
from several days to weeks. The presence of superhumps in general indicates low
mass ratios and dynamically perturbed accretion disks, making them
a valuable tracer for identifying system properties and evolutionary
states \citep{kato2013,Joshi2025AandA}. Clear distinctions
between classes are not always evident due to transitional phases,
hybrid behaviour, and observational limitations \citep{zorotovic2016}.
Some systems display changes in behaviour over time or exhibit
properties from multiple classes, complicating their classification
\citep{woudt2002, rodriguez2007}.

The long-term evolution of CVs is fundamentally governed by angular momentum loss (e.g. \citet{Knigge2011}). In long-period systems ($P_{\mathrm{orb}} > 3$ hours), angular momentum loss is predominantly attributed to magnetic braking (MB) \citep{rappaport1983, spruit1983}. MB occurs when the differential rotation of the main sequence donor generates internal magnetic fields that couple with the stellar wind, carrying away angular momentum \citep{kawaler1988, reiners2008}. This slows the star's rotation and, consequently, the orbital evolution of the system \citep{verbunt1981}. In short-period systems ($P_{\mathrm{orb}} < 2$ hours), gravitational radiation (GR) is thought to be the dominant mechanism \citep{paczynski1967, faulkner1971}. 

Several key observational features characterize the orbital period distribution of
cataclysmic variables. One of the most prominent is the orbital period gap, defined
by the observed scarcity of systems with orbital periods between approximately
2–3 hours \citep{patterson1998, Knigge2011}. Another important feature is the
existence of a minimum orbital period, below which CVs are not observed to evolve.
Systems that have passed this minimum and subsequently evolved back toward
longer orbital periods are known as period bouncers, and are expected to accumulate
near the period minimum \citep{gansicke2009, Knigge2011, munoz2024}.

These observational characteristics form the basis of the standard model of CV
evolution, which aims to explain the structure of the orbital period distribution.
As CVs evolve from longer to shorter orbital periods, angular momentum loss is
initially dominated by magnetic braking. This mechanism is believed to cease
abruptly when the donor star becomes fully convective at an orbital period of
approximately 3 hours \citep{spruit1983, schreiber2024}. The sudden reduction in
angular momentum loss causes mass transfer to halt temporarily, leading to the
detachment of the binary system.

Following detachment, the system continues to evolve toward shorter orbital
periods through gravitational radiation alone. When the orbital period reaches
approximately 2 hours, the donor star once again fills its Roche lobe and mass
transfer resumes \citep{Knigge2011, schreiber2024}. As evolution proceeds, the
donor becomes increasingly degenerate and is no longer able to contract in
response to mass loss. Consequently, its radius begins to increase, reversing the
direction of orbital evolution and producing the population of period bouncers
observed near the minimum orbital period ($P_{\mathrm{min}}$)
\citep{patterson1998}.

However, several inconsistencies between this model and observational data remain unresolved \citep{schreiber2024, sarkar2024}. First, the observed minimum period ($\sim$82 min)\citep{gansicke2009, mcallister2019} is significantly longer than the theoretically predicted value ($\sim$65 min) \citep{Kolb1999}, likely due to thermal disequilibrium and radius inflation in the donor star. This discrepancy is thought to arise from angular momentum loss in excess of that expected from pure gravitational radiation, which leads to enhanced mass-transfer rates and drives the donor out of thermal equilibrium. Second, the standard model predicts that the majority of present-day CVs should be short-period systems and period bouncers; yet, observational surveys do not reflect this distribution \citep{munoz2024, pala2020, schreiber2024}. Third, the assumption that magnetic braking ceases entirely when the donor becomes fully convective has been questioned \citep{chabrier2006, reiners2008, morin2010, garraffo2018, belloni2023, sarkar2024}. Recent studies suggest that fully convective low-mass stars---and even brown dwarfs---may sustain magnetic activity and continue to experience spin-down via MB-like processes, albeit with lower efficiency \citep{chabrier2006,reiners2008, morin2010, garraffo2018, belloni2023, sarkar2024}. Thus, the complete cessation of MB may not occur, presenting a potential conflict within the standard model \citep{schreiber2024}.

Although broader studies have been conducted to address these issues, the problems have not yet been fully resolved \citep{Knigge2011, pala2020}. Understanding the evolution of CV systems requires high-precision, long-term observational data across a wide sample \citep{pretorius2008_hALPHA, pala2020}. Detailed period analyses of known CVs, re-evaluation of misclassified systems, and discovery of new objects can all contribute to refining the standard evolutionary framework. Future gravitational wave missions, such as The Laser Interferometer Space Antenna (\textit{LISA}) \citep{amaroseoane2017laserinterferometerspaceantenna}, may also play a key role in determining the true location of the period minimum by directly probing the orbital dynamics of compact binaries \citep{Scaringi2023_CV_LISA}.

In this context, large-scale and high-confidence catalogues of cataclysmic variable (CV) stars are of great importance for understanding the evolutionary processes of binary star systems. By integrating observational data from heterogeneous sources into a consistent framework, such catalogues play a critical role in bridging the gap between theoretical predictions and observed system properties \citep{Goliasch_2015, schreiber2016}. Catalogues with high-confidence classifications \citep{ritter2003catalogue, Downes2001} provide comprehensive information on orbital periods, photometric characteristics, and fundamental system parameters. Since their initial publications, these catalogues were updated for a number of years\footnote{\url{https://archive.stsci.edu/prepds/cvcat/}
; \url{https://wwwmpa.mpa-garching.mpg.de/RKcat/}.}. 



In this paper, we aim to contribute to ongoing efforts in the field by presenting a new catalogue based on the systematic analysis of 2544 CV and CV candidate sources. The high time resolution and long-baseline photometric observations provided by the \textit{\textit{TESS}} mission enable a homogeneous and systematic evaluation of orbital periods in CV systems. The period values presented in this catalogue offer the opportunity to verify those reported in existing catalogues or to propose revisions where inconsistencies are suspected. 

To place our results in context and facilitate comparison with previously studied systems, we further cross-matched all analysed sources against the Ritter \& Kolb catalogue of cataclysmic variables \citep{ritter2003catalogue}. We also compared our results with several recent catalogues based on \textit{TESS} observations, allowing us to assess the consistency of the detected orbital periods and source
classifications across independent studies. In addition, we provide the orbital period distribution of these sources, offering new insights into the global properties of the CV population. The resulting catalogue aims to contribute to a more systematic classification of cataclysmic variable systems and provide a valuable reference for future studies involving detailed timing analysis.

The remainder of this paper is organised as follows. In Section~\ref{sec:data}, we describe the data selection and preparation process, including details of the \textit{TESS} observations. Section~\ref{sec:method} outlines our analysis techniques, such as period search methods and classification criteria. In Section~\ref{sec:results} we present the results of our frequency analysis together with a discussion of their implications for CV evolution. Finally, Section~\ref{sec:conclusion} summarises our findings and outlines prospects for future work. 

\section{DATA}
\label{sec:data}

\subsection{TESS Mission and Data Characteristics}
The primary dataset used in this study was obtained from \textit{TESS}, which has become a key resource for the detection and classification of cataclysmic variable stars (CVs) \citep{bruch2022, Littlefield2023SON, green2023, Bruch2024, bruch2024_SUPERHUMP, PichardoMarcano2021, Hernandez_2025}. \textit{TESS} delivers nearly continuous light curves for each observing sector, with each sector lasting approximately 27 days. The mission provides a 20 second and a 2-minute cadence for a subset of pre-selected objects, as well as Full Frame Images (FFIs) of the CCDs every 30 minutes during the primary mission and every 10 minutes during the extended mission. These data enable the identification of short-timescale photometric variability that is characteristic of CV systems \citep{ricker2015, tesshandbook}. Moreover, the ability to combine data from multiple contiguous sectors provides long temporal baselines, which are essential for tracking periodic signals and investigating long-term variability \citep{scaringi22a, scaringi22b, bruch2022}. With its wide spectral response (6000--10,000~\AA), \textit{TESS} is particularly sensitive to faint and subtle variations in optical and near-infrared brightness, making it highly suitable for CV studies and complementary to previous missions such as \textit{Kepler} \citep{sullivan2015, Littlefield2023SON}.

\subsection{Sample Selection}
The CV sources and related candidates analysed in this study were selected using publicly available data obtained through various Guest Investigator (GI) programs conducted during \textit{TESS} observation Cycles~1 to~6 which cover both hemispheres. The GI Programs are designed to broaden the scientific scope of the \textit{TESS} mission by enabling researchers to pursue their own scientific objectives \footnote{\url{https://heasarc.gsfc.nasa.gov/docs/tess/proposing-investigations.html}}.
All data are publicly available through the Barbara A. Mikulski Archive for Space Telescopes (MAST).\footnote{\url{https://mast.stsci.edu}}. 

\begin{table*}
\centering
\caption{List of TESS GI programs and their respective Principal Investigators (PIs) for the CVs analysed in this study. Multiple proposals by the same PI are grouped.}
\label{tab:gi_programs}
\begin{tabular}{ll}
\hline
\textbf{GI Proposal ID(s)} & \textbf{Principal Investigator} \\
\hline
G011268, G022071, G03071, G03044, G04046, G05094, G06027 & Scaringi \\
G022237, G03180, G04165, G05135, G06152 & Rivera Sandoval \\
G022230, G03245, G04208 & Littlefield \\
G011123, G022126, G04152 & Schlegel \\
G022116, G03240 & Wood \\
G011235 & Garnavich \\
G022254 & Sion \\
G03284 & Schwab \\
G04009 & Ramsay \\
\hline
\end{tabular}
\end{table*}

To compile an initial list of CVs and CV candidates we began by compiling all targets from GI programs listed in Table \ref{tab:gi_programs}. THE programs are those that have specifically targeted CVs or related systems. By definition the sample is heterogeneous as different principal investigators (PIs) selected specific targets based on theor specific science goals. We can comment on the creation of the target lists with PI Scaringi (G011268, G022071, G03071, G03044, G04046, G05094, G06027), which have compiled a list of known CVs from the literature, as well as populating the target list with transients suspected to be CVs either from ASAS-SN, ZTF, or \textit{Gaia} alerts. The initial list had also been filtered to exclude any target with a close and brighter neighbour within 30 arcseconds. This additional cut has been introduced to mitigate contamination in the lightcurve from a close coincident source. The initial compilation consisted of 2561 objects drawn from the GI programmes in Table \ref{tab:gi_programs}. We performed a SIMBAD cross -match to remove known and documented systems that are not categorised as CVs or related objects. This step reduced the initial list to  2544 objects, where most of the targets removed were known RR Lyrae-type stars.

Several observational biases must be considered in the target selection and period detection process using TESS data. One of the most significant limitations arises from the large pixel scale of the TESS detectors (approximately 21 arcseconds per pixel). This wide pixel size increases the likelihood of flux contamination from nearby stars, particularly in crowded stellar fields. Such contamination can dilute the intrinsic variability of the target, potentially rendering periodic signals undetectable or introducing spurious periodicities originating from neighboring sources. To minimize such contamination, the Guest Investigator programs with PI Scaringi (G011268, G022071, G03071, G03044, G04046, G05094, G06027) tend to select isolated sources without nearby stars, introducing a selection bias in favour of uncrowded regions.

TESS photometry becomes increasingly noise-dominated for sources fainter than ~17–18 mag \citep{Stassun2018, sullivan2015, Hernandez_2025}, which makes the detection of low-amplitude variability more challenging. In this study, the mean G-band magnitude of the analysed sources is found to be 16.76 mag, further indicating a bias toward intrinsically brighter systems. Moreover, \textit{TESS}’s observing strategy leads to non-uniform sky coverage: regions near the ecliptic poles are observed continuously or nearly continuously across many sectors, whereas those near the ecliptic plane are typically observed only once or twice per year. This uneven temporal coverage reduces the likelihood of detecting periodic signals in systems with low-amplitude or intermittent variability. These factors must be taken into account when assessing the completeness and representativeness of the resulting catalogue.

\section{METHODOLOGY}
\label{sec:method}

Orbital period searches in cataclysmic variables are most commonly performed using Lomb--Scargle periodograms \citep{Scargle1982} and their generalised forms \citep{Zechmeister2009}, often complemented by phase-folding techniques such as analysis-of-variance (AoV; \citealt{SchwarzenbergCzerny1989}), phase-dispersion minimisation (PDM; \citealt{Stellingwerf1978}), autocorrelation functions \citep{Edelson1988}, and Fourier power spectra \citep{Deeming1975} (e.g. \citealt{VanderPlas2018}; \citealt{Hernandez_2025}).
Orbital periods in unevenly sampled photometric time series are most
commonly identified using Lomb--Scargle periodograms and their
variants \citep{Lomb1976,Scargle1982}. This framework has been widely
adopted in recent \textit{TESS} studies of cataclysmic variables (CVs)
\citep{bruch2022,Littlefield2023SON,Hernandez_2025}. However,
CV light curves are frequently dominated by flickering and correlated (red) noise, which leads to a frequency-dependent noise floor in periodograms. Under such conditions, the use of a single global detection threshold can either suppress genuine signals in high-noise regions or inflate the number of false positives in quieter parts of the spectrum.

To mitigate this effect, we adopt a window-based detection scheme in
which the significance threshold is estimated locally within sliding
frequency windows. In this approach, robust statistics are computed
after trimming the central portion of each window to reduce the
influence of strong peaks and localised artefacts. This methodology
is conceptually similar to robust background estimation techniques
commonly employed in power-spectrum analyses of red-noise-dominated
time series, but is here applied systematically to large samples of
CVs. By adapting the detection threshold to the local noise
properties of the periodogram, our method aims to maintain
conservative control over false detections while improving
sensitivity to weak but coherent signals. This is particularly
advantageous for identifying orbital, spin and superhump
modulations in cataclysmic variable systems.


The input list includes the \textit{TESS} Input Catalog Identifier
(TIC ID), a unique numerical designation assigned to each object in the
\textit{TESS} Input Catalog \citep{Stassun2018,
Stassun2019_RevisedTIC_CTL}. For each target, all \textit{TESS} sectors
with available data were identified, and the corresponding light curves
were retrieved. \textit{TESS} photometric data are provided through the
MAST archive in two forms: Simple Aperture Photometry (SAP) and
Pre-search Data Conditioning Simple Aperture Photometry (PDCSAP).
SAP light curves consist of minimally processed flux measurements
extracted directly from a predefined photometric aperture, preserving
the intrinsic astrophysical variability of the source. In contrast,
PDCSAP light curves are produced by applying an additional conditioning
pipeline designed to mitigate common instrumental systematics and
long-term trends.
For each sector associated with a given source, flux measurements were
extracted using the SAP light curves (\citealt{Jenkins2016_TESS_SPOC},
SAP\_FLUX). We adopt SAP data in order to preserve intrinsic
astrophysical variability, which is particularly important for orbital
period determination in cataclysmic variable systems. While PDCSAP
processing is effective at reducing instrumental effects, the
conditioning procedure can in some cases distort genuine astrophysical
variability in cataclysmic variable light curves \citep{bruch2022}.
Only data points with a quality flag equal to zero, indicating the
absence of known instrumental or observational issues, were included in the analysis.

After this cleaning step, the Lomb--Scargle method
\citep{Lomb1976,Scargle1982} was applied to compute the periodogram for
each sector of each target. All light curves were analysed on a
sector-by-sector basis and no multi-sector light curves were used.
The frequency grid extended up to $360~\mathrm{cycles~day^{-1}}$, which
corresponds to the Nyquist frequency for 2-minute cadence data, ensuring
sensitivity to short-period modulations without introducing aliasing
effects. The lower bound was set to 1 cycle/day instead of the minimum
frequency defined by the total time span of the observations to avoid
spurious low-frequency signals from long-term systematics and red-noise
broadband variability typical of CVs \citep{scaringi14}, focusing the
automated search on coherent periodicities of astrophysical interest,
e.g., orbital and spin/beat periods. Nevertheless, during manual
inspection, several sources exhibited coherent long-period modulations
below 1 cycle/day. These cases were individually flagged and included
in the final catalogue when the signal appeared significant (see
Section~\ref{sec:method}).

\subsection{Frequency Detection}

The frequency detection algorithm is based on two tunable parameters: the window size and the trimming fraction. Windows are defined in terms of frequency bins, each bin corresponding to a fixed step size of $\Delta f = 0.007~\mathrm{cycle~d^{-1}}$. This step size was chosen based on the intrinsic frequency resolution of the data. For a time series with a total observational baseline $T$, the Rayleigh frequency resolution is given by $\Delta f_{\mathrm{R}} \simeq 1/T$. For a typical TESS sector with $T \approx 27$~days, this corresponds to $\Delta f_{\mathrm{R}} \approx 0.037~\mathrm{cycle~d^{-1}}$. The adopted step therefore oversamples the Rayleigh resolution by a factor of $\sim5$, ensuring that narrow peaks in the periodogram are adequately resolved.

The number of bins in a window determines the effective frequency resolution within each segment; for example, a window containing 50 bins spans approximately $0.35~\mathrm{cycle~d^{-1}}$. After testing a range of parameter combinations, a window size of 50 bins with a central trimming of 20 bins (40\% trimming) was found to provide the optimal balance between sensitivity to weak signals and minimization of false positives. This frequency grid provides sufficient resolution to detect short-period signals on timescales of minutes to hours across the full frequency range considered. For instance, at the upper end of our search range ($f \sim 300~\mathrm{cycle~d^{-1}}$, corresponding to a period of $\sim4.8$~minutes), the adopted step samples each peak with multiple grid points, preventing the loss of high-frequency periodicities due to coarse frequency sampling.

Within each trimmed window, a dynamic detection threshold was computed as

\begin{equation}
\hspace*{1cm}
\text{Threshold}(f_i) = \mu_{\text{trimmed}}(f_i) + 10 \times \sigma_{\text{trimmed}}(f_i) ,
\label{eq:threshold}
\end{equation}

\noindent where $\mu_{\rm trimmed}(f_i)$ and $\sigma_{\rm trimmed}(f_i)$ denote the mean and standard deviation of the trimmed power spectrum values within the window centred at frequency $f_i$. The trimming procedure removes the central bins of the window, which typically contain the highest-power values when a real signal is present, thereby preventing the signal itself from biasing the local noise statistics.

As the sliding window approaches a real peak from either side, elevated power in the wings of the peak increases both the local mean and standard deviation, leading to a raised detection threshold. In contrast, when the window is centred on the peak, the exclusion of the
central bins results in a locally reduced mean and standard deviation, producing a minimum in the detection threshold and enabling robust
detection of the signal at its true frequency.

The threshold multiplier $k = 10$ was selected after testing a range of values, ensuring a conservative detection limit that suppresses
spurious detections near strong peaks and in regions of elevated noise. Within each window, frequencies exceeding the local threshold were
flagged as significant detections. After scanning the entire periodogram, the lowest among these significant frequencies was selected as the first fundamental frequency ($f_{\mathrm{fund}}$), which
typically corresponds to the orbital frequency of the system.

To reliably identify the second fundamental frequency, the algorithm includes a harmonic identification step. Harmonics are defined as integer multiples of the fundamental frequency, as expressed in Equation~\ref{eq:harmonic}:

\begin{equation}
\hspace*{2.3cm}
f_{n} = n\, f_{\rm fund}, \qquad n \in \mathbb{Z}^{+},
\label{eq:harmonic}
\end{equation}

In real data, harmonics do not satisfy this identity exactly, but instead deviate slightly from the ideal integer-multiple relation. The measured frequency can therefore be written as

\begin{equation}
\hspace*{1.8cm}
\hat{f}_{n} = n\, f_{\rm fund} + \delta_{n},
\qquad |\delta_{n}| \ll n\, f_{\rm fund},
\label{eq:realistic_harmonic}
\end{equation}

A frequency is classified as a harmonic if it satisfies the following practical condition:

\begin{equation}
\hspace*{2.3cm}
\left| \hat{f}_{n} - n\, f_{\rm fund} \right| < \varepsilon ,
\label{eq:epsilon}
\end{equation}

\noindent where the tolerance parameter $\varepsilon = 0.1$. This value was chosen empirically to balance harmonic identification with the frequency resolution of the periodogram, avoiding both missed harmonics and excessive inclusivity.

After all harmonics of the first fundamental frequency were identified, the next lowest frequency not classified as a harmonic was recorded as the second fundamental frequency. In CV systems, this second frequency is typically associated with the spin period of the WD or a superhump signal. All detected frequencies were recorded and subjected to harmonic analysis, and the algorithm not only retains the first and second fundamentals but, in fact, records all apparent fundamental signals, attempting to classify them according to their harmonic relationships. Also note, the term ``first harmonic'' refers to the component at $2f_{\rm fund}$, as the fundamental frequency in this study.

\begin{figure*}
    \centering
    \includegraphics[width=0.90\textwidth]{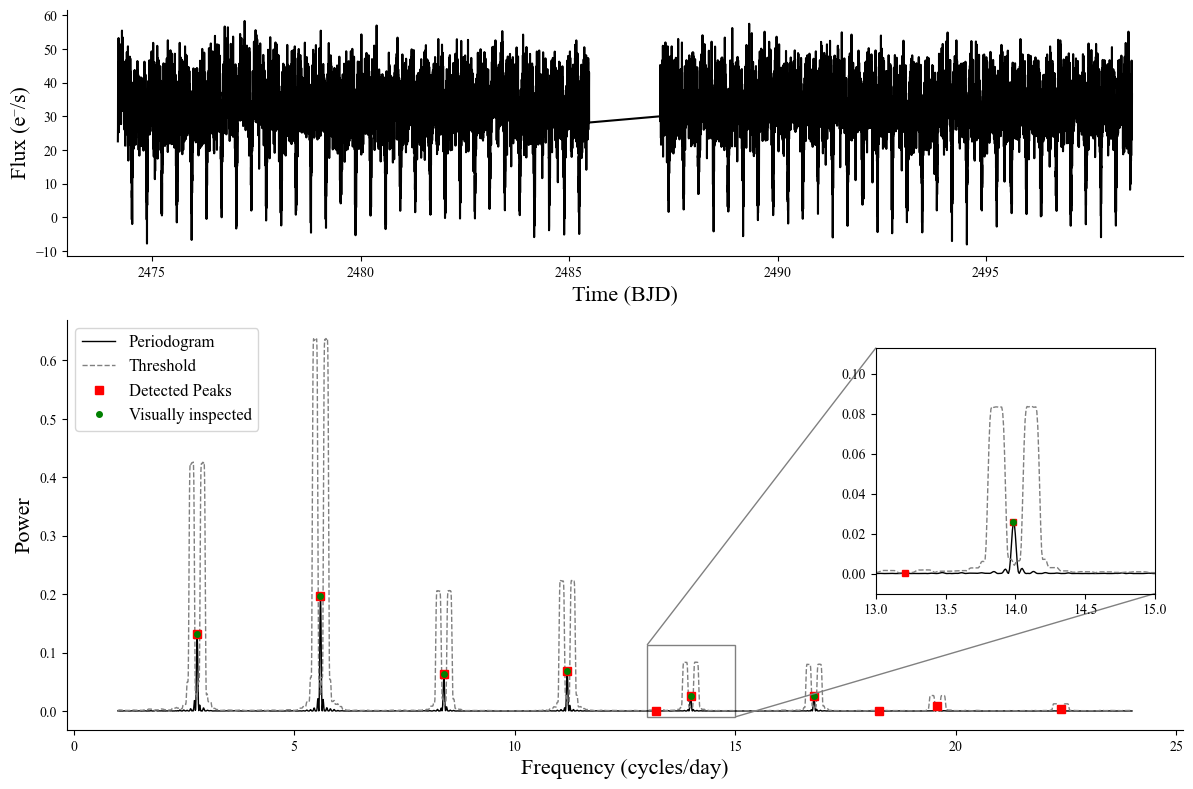}
    \caption{\textit{TESS} Sector 43 light curve (upper panel) and Lomb–Scargle periodogram (lower panel) of the cataclysmic variable V* QZ Aur (TIC ID  3034524). The threshold, defined as $\mu + k\sigma$ with $k = 10$, is shown in grey. All peaks above this threshold are marked in red, while visually confirmed signals are highlighted in dark green. The inset provides a zoomed-in view frequency near 14 cycles day$^{-1}$ to better illustrate the peak structure and detection method.}
    \label{fig:TIC ID _smooth}
\end{figure*}

Figure~\ref{fig:TIC ID _smooth} provides an example output of the periodogram analysis applied to the source \textit{V* QZ Aur} (TIC ID  3034524) using the method described in this study. The upper panel shows the light curve of the source as observed by \textit{TESS}. The lower panel presents the application of the algorithmically determined threshold to the periodogram, with the frequencies exceeding the threshold clearly marked. The peaks detected by the algorithm are marked in red, while those that were subsequently confirmed as significant through visual inspection are highlighted in green on top of the red points. Signals that remain in red were judged to be noise-dominated and not intrinsic to the system, and were therefore left in red to distinguish them clearly from the significant detections.

This methodology was applied to the full dataset comprising 2544 sources. The initial output includes all sources for which the algorithm identified at least one significant frequency. Each entry contains the TIC ID   of the source, the \textit{TESS} sector in which the frequency was detected, the first fundamental frequency, and its harmonics, if present. In cases where a second fundamental frequency was found, it is also reported. All frequencies are expressed in cycles per day (cycle d$^{-1}$).

\begin{table*}
\centering
\renewcommand{\arraystretch}{1.2}
\caption{
This table presents a representative subset of five systems from the CV Confident Catalogue.
From left to right, the columns provide the TESS Input Catalog identifier (TIC ID ), the commonly adopted source name, and the TESS sectors in which the light curve was observed.
The first and second fundamental frequencies, expressed in cycles per day (c d$^{-1}$), correspond to the final values obtained after the bootstrap-based uncertainty analysis.
The next two columns indicate the presence of negative (nSH)  or positive superhumps (pSH), marked with ‘Y’ where detected. 
The Extra Info and Appendix columns include additional information from the literature and specify whether a detailed figure of the source is provided in the Appendix.
The final column is the orbital frequency of the system ($f_{\rm orb}$ (cycle d$^{-1}$)) which is listed only when the orbital frequency can be trusted based on the fundamental frequency identified by the algorithm.
}

\label{tab:random10_cv}
\resizebox{\textwidth}{!}{%
\begin{tabular}{ccccccccccccc}
\toprule
\textbf{TIC ID } & \textbf{Name} & \textbf{LC TESS Sectors} &
\textbf{1st fund (cycle d$^{-1}$)} & \textbf{2nd fund (cycle d$^{-1}$)} &
\textbf{nSH} & \textbf{pSH} &
\textbf{Extra info} & \textbf{Appendix} &
\textbf{$f_{\rm orb}$ (cycle d$^{-1}$)} &
\textbf{RA} & \textbf{Dec} \\
\midrule
2028705173  & Cl* NGC 7099 SAW V4 & 68         & 11.425  & 18.976  & -- & -- & -- & -- & 11.425 & 324.9935 & -23.1955 \\
8765832     & V* BK Lyn           & 21         & 13.343  & --      & -- & Y  & -- & Y  & 13.243 & 140.0467 & 33.9451  \\
121992913   & V* V587 Lyr         & 41;54;80   & 3.650   & --      & -- & -- & -- & Y  & 3.650  & 289.3603 & 37.17801 \\
11116617    & RX J2015.6+3711     & 41;55;82   & 3.7625  & 4.5829  & -- & -- & -- & -- & --     & 303.9040 & 37.1897  \\
903265195   & ASASSN-17fo         & 72         & 16.240  & --      & -- & -- & -- & -- & 16.240 & 174.6487 & 4.748528 \\
\bottomrule
\end{tabular}}%
\end{table*}

\subsection{Bootstrap Error Estimation}

The uncertainties of the first and second fundamental frequencies for the entire catalogue were determined using a bootstrapping method
\citep{Bootstrap_first, 10.1093/mnrasl/slae035, Irving2024, Veresvarska2025_DW_Cnc}.
For each source--sector pair, a bootstrap procedure with 10000 iterations was performed by resampling the original light curve with replacement, drawing \(N\) data points, where \(N\) equals the length of the original dataset.
For each iteration, a Lomb--Scargle periodogram was computed within a frequency window of
\(\pm 0.2~\mathrm{cycle~d}^{-1}\) around the target fundamental frequency using a grid of 5{,}000 frequency points, and the frequency corresponding to the maximum power was recorded.

\begin{figure}
\centering
\begin{minipage}{1\linewidth}
    \centering
    \includegraphics[width=\linewidth]{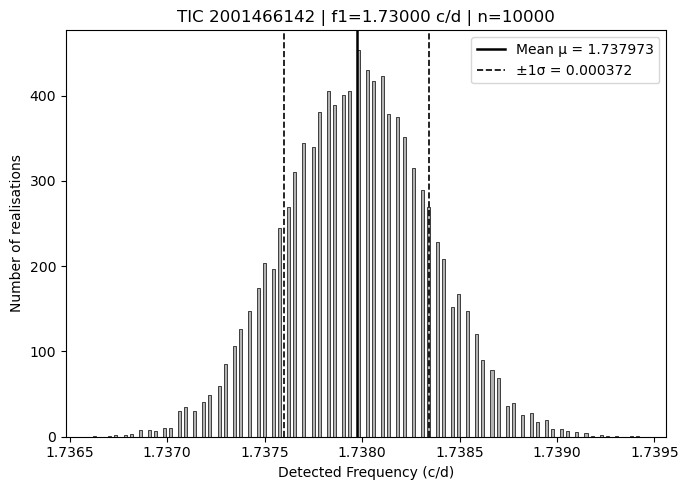}
    \caption{The histogram shows the distribution of recovered peak frequencies obtained from 10000 bootstrap realisations. The solid vertical line marks the bootstrap mean frequency ($\mu = 1.73797~\mathrm{c/d}$), while the dashed vertical lines indicate the $\pm1\sigma$ uncertainty ($\sigma = 0.00037~\mathrm{c/d}$). The y-axis represents the number of realisations.}
    \label{fig:error_analysis_tic2001466142}
\end{minipage}
\end{figure}

The distribution of these maximum--power frequencies was constructed for each sector.
The uncertainty was estimated directly from the 16th and 84th percentiles of the bootstrap distribution, corresponding to the \(\pm1\sigma\) confidence interval, while the median of the distribution was adopted as the best estimate of the signal frequency.
An example bootstrap distribution for ASASSN--14cl is shown in Fig.~\ref{fig:error_analysis_tic2001466142}.

The final frequency value for each source was then calculated as the inverse--variance weighted mean of the sector measurements.
Given the bootstrap median frequency \(\mu_i\) and its associated uncertainty \(\sigma_i\) for each sector, the relation in Equation~\ref{eq:error} was applied.

\begin{equation}
\hspace*{1.5cm}
\mu_{\mathrm{final}}
=\frac{\displaystyle\sum_{i}\frac{\mu_i}{\sigma_i^{2}}}
{\displaystyle\sum_{i}\frac{1}{\sigma_i^{2}}},\qquad
\sigma_{\mathrm{final}}
=\left(\sum_{i}\frac{1}{\sigma_i^{2}}\right)^{-1/2}
\label{eq:error}
\end{equation}

Fig \ref{fig:error_analysis_tic2001466142} shows the distribution of the detected fundamental frequency ($f_{1}$) 
    from 10\,000 bootstrap realisations. The red curve is a Gaussian fit, yielding a mean 
    of $\mu = 1.73797~\mathrm{c\,d}^{-1}$ with a standard deviation of 
    $\sigma = 0.00037~\mathrm{c\,d}^{-1}$. Vertical dashed lines indicate the mean value 
    and the $\pm1\sigma$ confidence intervals.

\subsection{Contamination Analysis}

To evaluate possible contamination from nearby sources, we cross-matched
our catalogue with the Variable Star Index (VSX), maintained by the
American Association of Variable Star Observers (AAVSO), using a $60\arcsec$ search radius.The choice of VSX is motivated by our primary concern regarding contamination
by \textit{variable} stars. Given the large TESS pixel scale ($\sim21~\arcsec~\mathrm{pixel}^{-1}$), multiple nearby sources can fall within the same photometric aperture. While flux dilution from
non-variable neighbours primarily reduces the observed variability amplitude, it does not by itself introduce spurious periodicities. In
contrast, variability from nearby variable stars can lead to the detection of periodic signals that are not intrinsic to the target
source. By cross-matching with VSX, we therefore specifically identify cases in which detected periodicities may plausibly originate from known nearby variable objects rather than from the target itself.

A total of 132 targets were found to have neighbouring sources within 60~arcsec. For each of these systems, all objects falling inside this angular window were identified, and the number of such neighbouring sources and their angular separations were recorded. All such cases were then visually inspected to identify potential blending or misidentification, ensuring that the main target was correctly associated with the detected periodic signal.

Furthermore, for each VSX-matched group, the brightest neighbouring source (based on the $V$ magnitude) was identified and its brightness compared with that of the main target. Systems hosting a brighter neighbouring star than the main target were examined in greater detail. Sources confirmed to be contaminated, or showing a high risk of contamination but lacking sufficient classification or literature information for confirmation, were flagged in CCC. As a result, based on VSX data, 57 targets were found to have a brighter neighbouring variable star within 60~arcsec of the main object, indicating potential or confirmed contamination. 

Sources exhibiting definite contamination are flagged with C, while those with a possible contamination risk are marked as C? in the Extra info column. A table presenting five of these sources is shown in Table \ref{tab:contamination}. We chose not to remove some of the definitively contaminated sources from our final sample to ensure that the catalogue remains as complete and representative as possible.  Only a single object, OGLE MC-DN-30 (TIC ID 735229757), was excluded, as all detected signals in this case were found to originate from the contaminating neighbour rather than the main target. Further details for this system are provided in the corresponding entry in the Appendix. For the remaining definitively contaminated sources, at least one detected signal was still likely to originate from the main target, or an additional periodicity was present that plausibly represents the true orbital modulation of the system. For this reason, these sources were retained in the final sample, despite the presence of contaminating neighbours. 

\begin{table*}
\centering
\caption{Inset table showing five sources flagged as contaminated (C) or potentially contaminated (C?), selected as representative examples of systems affected by nearby stars whose brightness or variability may influence the detected frequencies.}
\label{tab:contamination}
\resizebox{\textwidth}{!}{%
\begin{tabular}{ccccccccccccc}
\toprule
\textbf{TIC ID} & \textbf{Name} & \textbf{LC TESS Sectors} &
\textbf{1st fund (cycle d$^{-1}$)} & \textbf{2nd fund (cycle d$^{-1}$)} &
\textbf{nSH} & \textbf{pSH} &
\textbf{Extra info} & \textbf{Appendix} &
\textbf{$f_{\rm orb}$ (cycle d$^{-1}$)} &
\textbf{RA} & \textbf{Dec} \\
\midrule

290769912 & V* TW Vul & 41;55;82 &
4.8649 & 12.49 &
-- & -- & C & Y & 4.8649 & 309.8937 & 27.26551 \\

72182461 & V* V373 Cen & 64 &
3.7535 & -- &
-- & -- & C? & Y & 3.7535 & 186.5212 & -45.8252 \\

375982881 & ZTF17aaburxr* & 77 &
2.048 & 14.522 &
-- & -- & C & Y & -- & 336.9403 & 68.72742 \\

403018318 & 1RXS J150618.6-750157* & 65;66 &
12.7481 & -- &
-- & -- & C? & Y & 12.7481 & 226.5666 & -75.0336 \\

464626077 & Gaia 19bdy & 63;64 &
7.237 & -- &
-- & -- & C & Y & 7.237841 & 156.8021 & -58.4352 \\

\bottomrule
\end{tabular}}%
\end{table*}

\section{Results and Discussion}
\label{sec:results}
The main objective of this paper is to present a catalogue compiled by analysing all sources proposed in the initial list, including only those that are believed to exhibit genuine signals as a result of this analysis. 

\begin{figure*}
    \centering        \includegraphics[width=0.8\textwidth]{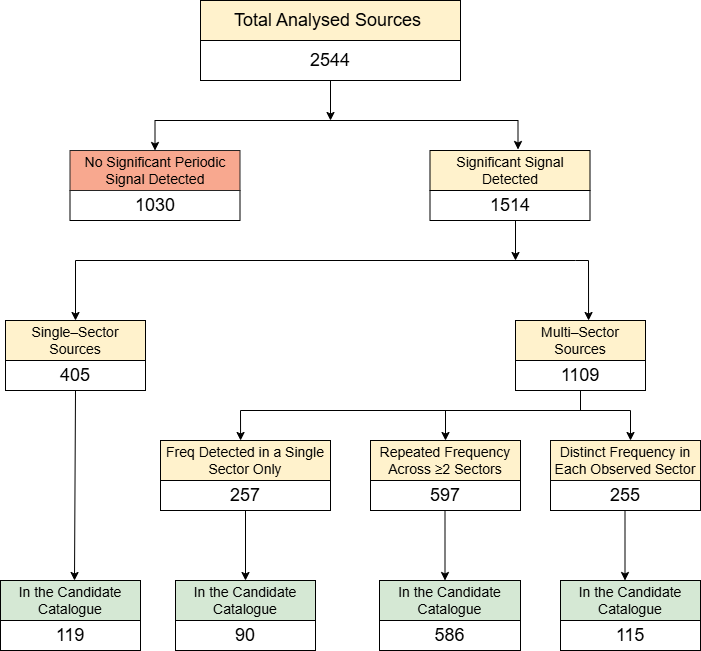} 
        \caption{Distribution of the analysed sources based on the number of observed sectors and the number of detected frequencies.All sources with green-labeled numbers have been included in the CV Confident Catalog.}
        \label{fig:flowchart}
\end{figure*}

The sources included in the catalogue are categorised based on the number of \textit{TESS} sectors in which they were observed and the frequency values detected across these sectors. Each source was evaluated within this categorisation framework and further subjected to multiple visual inspections to complete the analysis. To provide a clearer understanding of this process, a flowchart summarizing the analysis steps is presented in Fig \ref{fig:flowchart}.

The flowchart presented here illustrates the distribution of sources based on the number of observed sectors and the number of detected frequencies, from the initial stage of the analysis to the final classification. At the beginning of the analysis, a list comprising 2544 sources was considered. As a result of the algorithm's evaluation, significant periodic frequencies were detected in 1514 sources, while no meaningful signal was found in the remaining 1030. One of the key reasons for the importance of the algorithm is that it removes the necessity of performing a visual inspection on the entire sample of 2544 sources. The sources deemed to show significant signals were then classified based on the number of sectors in which they were observed and the distribution of detected frequencies across these sectors.

\subsection{Sources with single-sector TESS coverage}

Among these 1514 sources, 405 of them were observed in only one sector of \textit{TESS} during the time range in which the algorithm retrieved the light curves. Therefore, we could analyse only the 2-minute cadence light curve from that single sector, and any significant frequencies identified in these cases could not be compared with other sectors.

Based on the signal-to-noise ratio (SNR) and visual inspection, orbital periods believed to be reliable were included in the list. However, only 119 out of the 412 sources passed this additional scrutiny. In this context, ``passing'' refers to cases where the detected signal exhibited a sufficiently high SNR relative to the surrounding periodogram noise, allowing for confident classification as a genuine periodic feature.  For sources with significant frequencies detected in only a single sector, inclusion in the final catalogue was based on a combination of criteria: the significance of the detected signal, the noise level in the periodogram, and the presence of harmonics associated with the recorded first fundamental frequency. Sources that were excluded generally failed to meet these criteria, with the primary reason being that the detected frequencies appeared to have power levels close to the mean signal level in the periodogram, making it difficult to confidently distinguish them from statistical fluctuations. All exclusion decisions were made on a case-by-case basis after individual visual inspections and required confirmation by at least two reviewers.

\begin{figure}
    \centering
    \hspace*{-0.8cm}
    \includegraphics[width=0.5\textwidth]{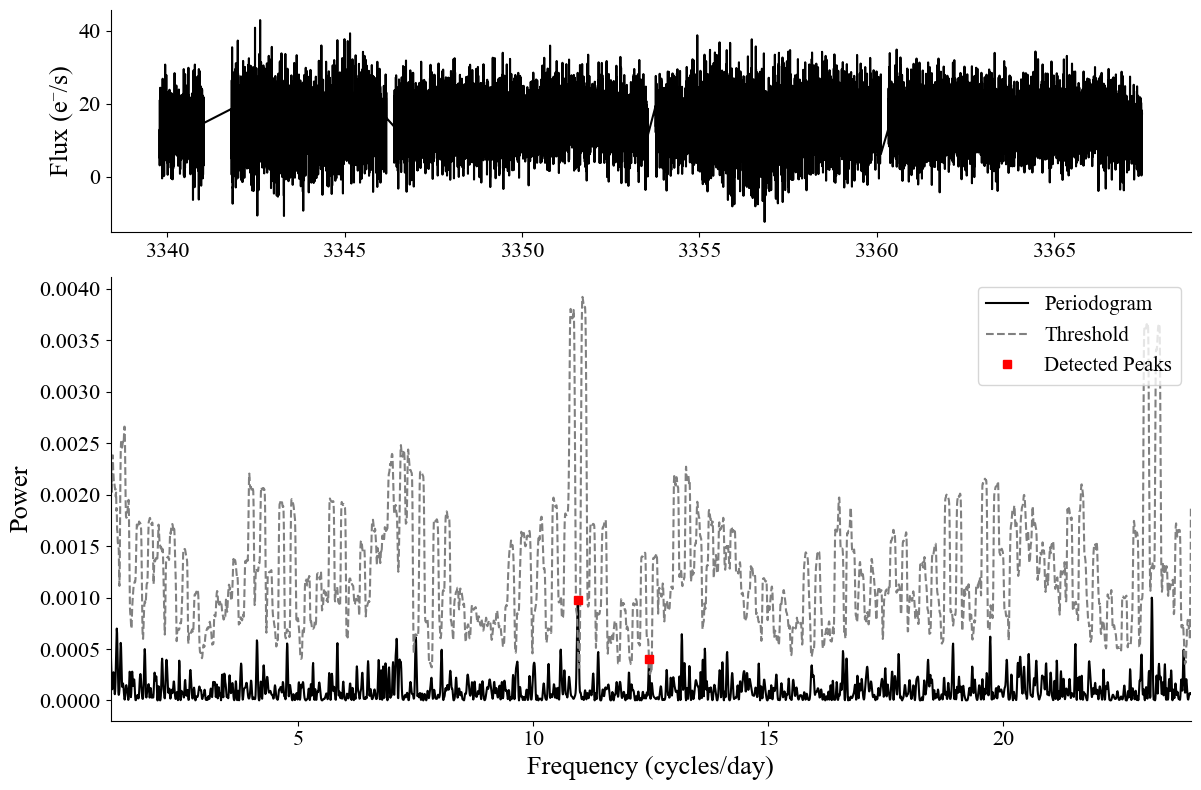}
    \caption{Top panel shows the TESS light curve of ZTF19abagxei (TIC ID  122519668) observed in Sector 75. Bottom panel presents the corresponding Lomb–Scargle periodogram, where the peaks at 10.94 and $\sim$13 cycles d$^{-1}$ (marked in red) are likely spurious due to the low power level and the lack of harmonic structure.}
    \label{fig:one_Sector_example}
\end{figure}

An illustrative example is provided in Fig. \ref{fig:one_Sector_example}, where the CV candidate source ZTF19abagxei \citep{ZTF_CV_Cands} is shown. The algorithm recorded a signal at 10.94 cycle d$^{-1}$; however, visual inspection suggests that the light curve might be substantially contaminated by noise. Additional data are needed to determine whether this signal is due to noise or a true periodicity.

The next subset comprises sources observed in multiple \textit{TESS} sectors, totalling 1109. However, evaluating periodic signals across multiple sectors introduces a layer of complexity. To address this, these sources were subdivided into three distinct categories. 

\subsection{Single-sector detections}

The first category includes sources that, despite being observed in multiple sectors, exhibited a significant frequency in only one of them. This subset contained 257 sources. Similar to the single-sector group, these sources were initially considered unreliable. The fact that a significant frequency was detected in only one sector, despite the source being observed in multiple sectors, further supports this interpretation. However, each was visually inspected to determine whether the signal might still be considered meaningful. As expected, the majority did not pass this stage, with 167 out of 257 deemed unreliable and excluded from the list. Among the reviewed sources, some were identified where the lack of detection in other sectors was likely due to a low SNR. 

\begin{figure}
    \centering
    \hspace*{-0.8cm}
    \includegraphics[width=0.5\textwidth]{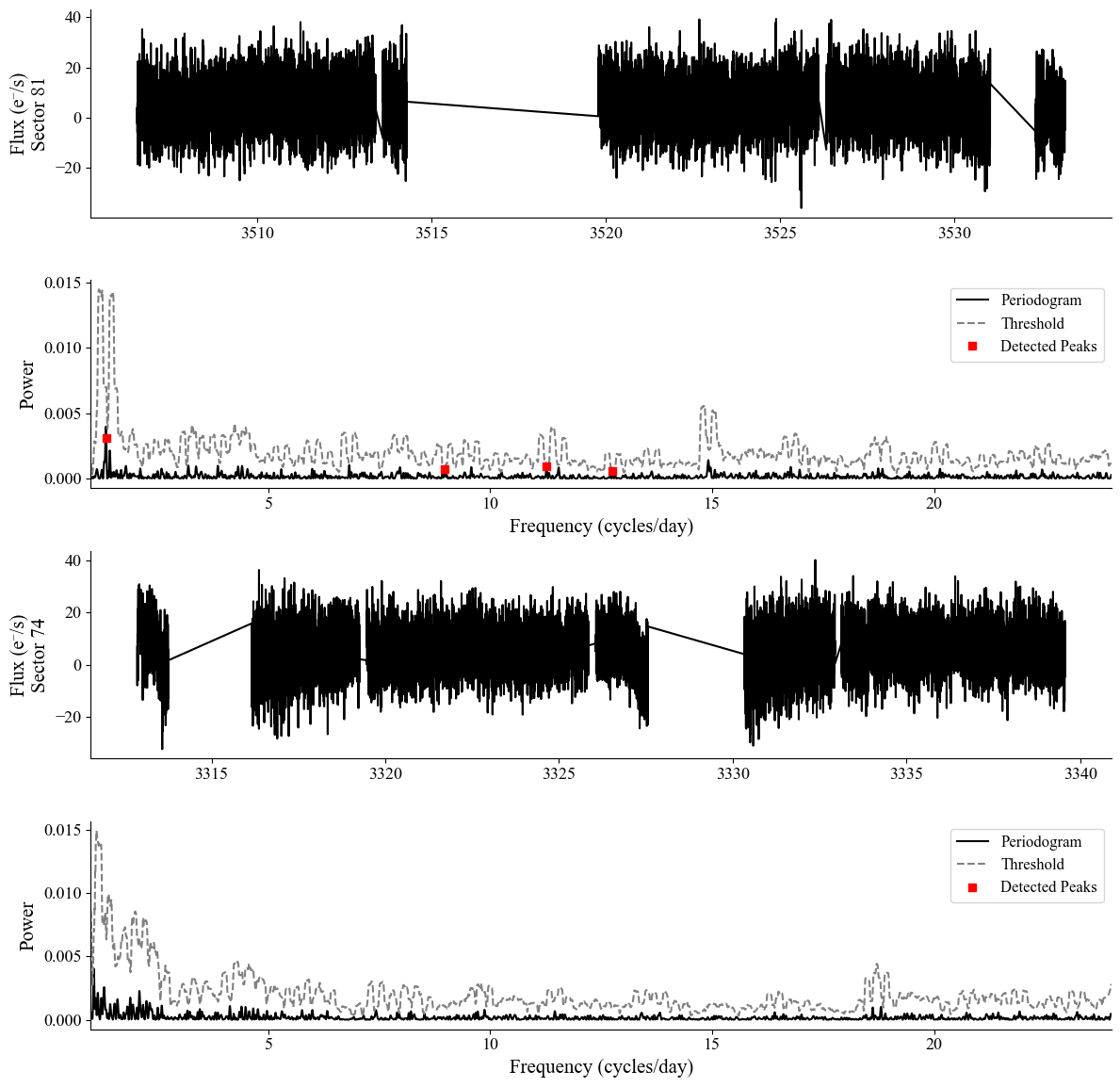}
    \caption{Light curves and periodograms of V2491 Cyg (TIC ID  1868684285) from Sector 81 (top panels) and Sector 74 (bottom panels). In Sector~81, the algorithm identified a peak at approximately 1.35~cycles\,day$^{-1}$ and classified it as significant (marked in red). However, visual inspection does not support the reality of this signal. No corresponding peak is present at this frequency in Sector~74, and both sectors exhibit low power levels without clear harmonic structure, indicating that the detection is likely spurious rather than a genuine periodic feature.} 
    \label{fig:multi_Sector_one_detection}
\end{figure}

An example of this is shown in Fig \ref{fig:multi_Sector_one_detection}. V2491 Cyg has three \textit{TESS} long cadence observations from Sectors 81, 75, and 74. The algorithm identified a frequency of 1.35 cycles/day in Sector 81, which it classified as significant. However, upon visual inspection, there is no indication that this signal is significant, neither in Sector 81 nor in the other two sectors. For this reason, the source was excluded from the final catalogue. Two signals at 10.44 and 1.41~cycles\,day$^{-1}$ were reported by \citet{Baklanov2008} and \citet{Zemko2018}, but \citet{Schaefer2022} did not confirm these in the TESS data, leaving their identification uncertain. We also note that V2491 Cyg is a nova that erupted in 2008, and systems so shortly after eruption may not yet have reached a stabilized quiescent accretion geometry.


\subsection{Sources with frequencies confirmed across multiple sectors}

The second category within the multi-sector group includes sources where the same frequency was detected in at least two different sectors. This group contains 597 sources and represents the most reliable subset, as the repeated detection of the same periodic signal across independent sectors significantly increases confidence in its authenticity. Nevertheless, these sources were still subjected to visual inspection, and 11 of them were ultimately excluded for not meeting the reported criteria. 


In most cases, the primary reason is that, although a signal may be formally significant, it is often buried within the noise and cannot be confidently verified upon visual inspection. Furthermore, we do not entirely rule out the possibility that these signals may indeed correspond to the true orbital periods. Rather, a more detailed analysis is required, which lies beyond the scope of this work and will be the subject of future investigation. The list of these 11 sources is provided in the Appendix.

An example of such a case is shown in Fig.\ref{fig:2_SECTOR_at_least}, where the light curves and periodograms of Gaia DR2 3451117071149973120 (TIC ID  312474735), a CV candidate, are presented for Sectors 43 and 45. In both sectors, the algorithm detected a frequency of approximately 4.01 cycles/day; however, after visual inspection, this source was not included in the final catalogue. This decision was made due to the closeness of the signal strength to the local noise level, which prevents a reliable interpretation. Additionally, it is important to note that this exclusion does not imply that the signal is definitively spurious. The source has light curve data from six \textit{TESS} sectors (Sectors 43, 44, 45, 71, 72, and 73); yet, this particular frequency was only detected in two of them. We note that changes in modulation amplitude could also arise from variations in the mass-transfer rate, although investigating the long-term brightness evolution is beyond the scope of this work.

\begin{figure}
    \centering
    \hspace*{-0.8cm}
    \includegraphics[width=0.5\textwidth]{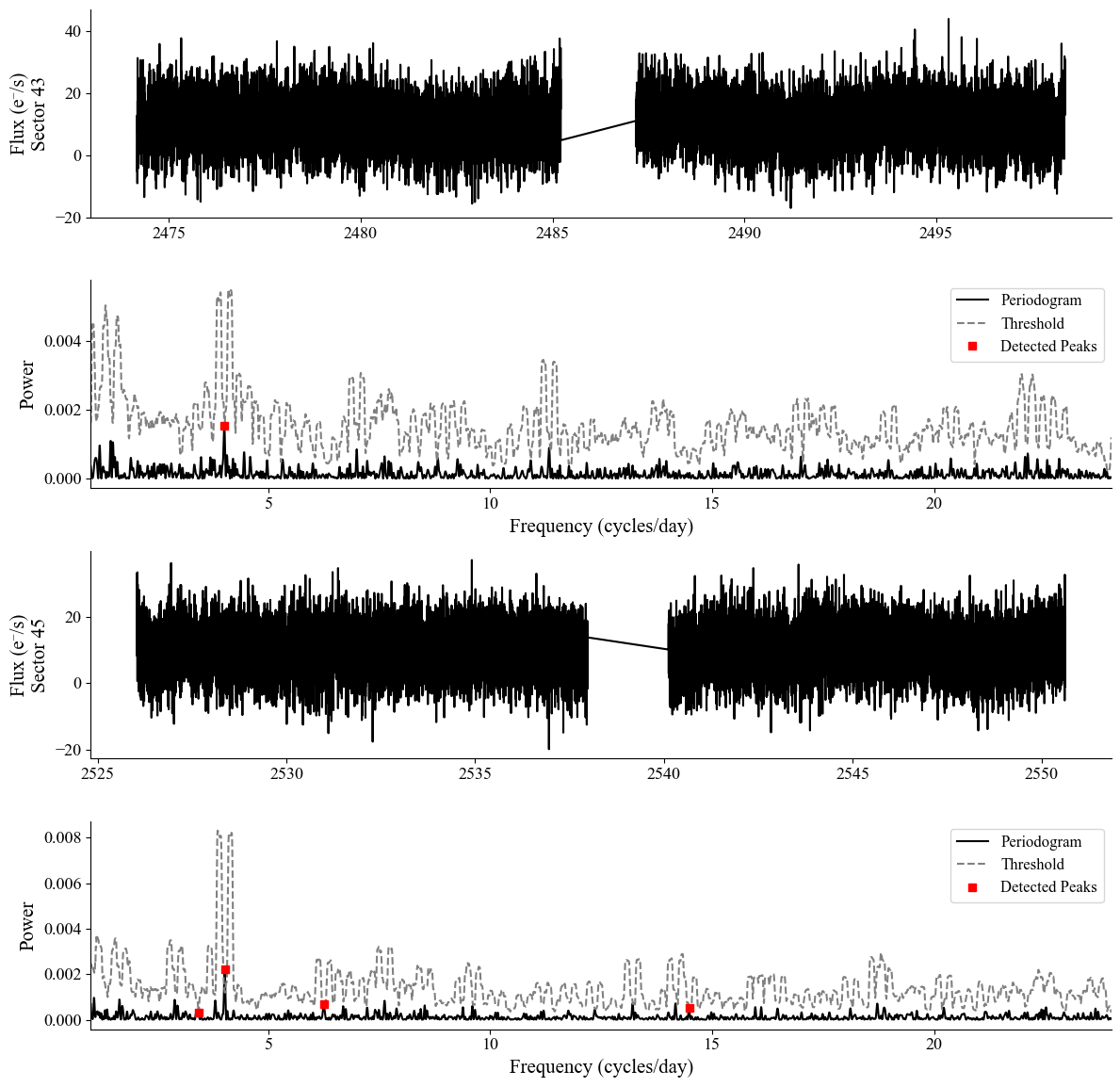}
    \caption{Light curves and periodograms of Gaia DR2 3451117071149973120 (TIC ID  312474753) observed in Sectors 43 (top panel) and 45 (bottom panel) are shown. In both sectors, the algorithm identifies a peak near 4.01 cycles d$^{-1}$, also flagged a few additional low-amplitude peaks 
    at other frequencies in sector 45, marked in red. Although the frequency at 4.01 cycles d$^{-1}$ is present in only 2 out of 6 sectors, the peak remains close to the noise level and insufficiently significant, and therefore the source was excluded from the final catalogue.} 
    \label{fig:2_SECTOR_at_least}
\end{figure}

\subsection{Sources with inconsistent multi-sector frequencies}

Another notable subcategory consists of sources where the detected frequencies differ between sectors, i.e. each sector contains a unique fundamental frequency. A total of 255 such sources were identified in this category. These cases were also carefully reviewed to determine their reliability. A recurring pattern observed during inspection was that such behaviour often corresponds to superhumping systems, where the first fundamental frequency may reflect usually a positive or negative superhump in some sectors, while in others, where the superhump disappears, the true orbital period is recovered as the fundamental frequency. 

\begin{figure}
    \centering
    \hspace*{-0.8cm}
    \includegraphics[width=0.5\textwidth]{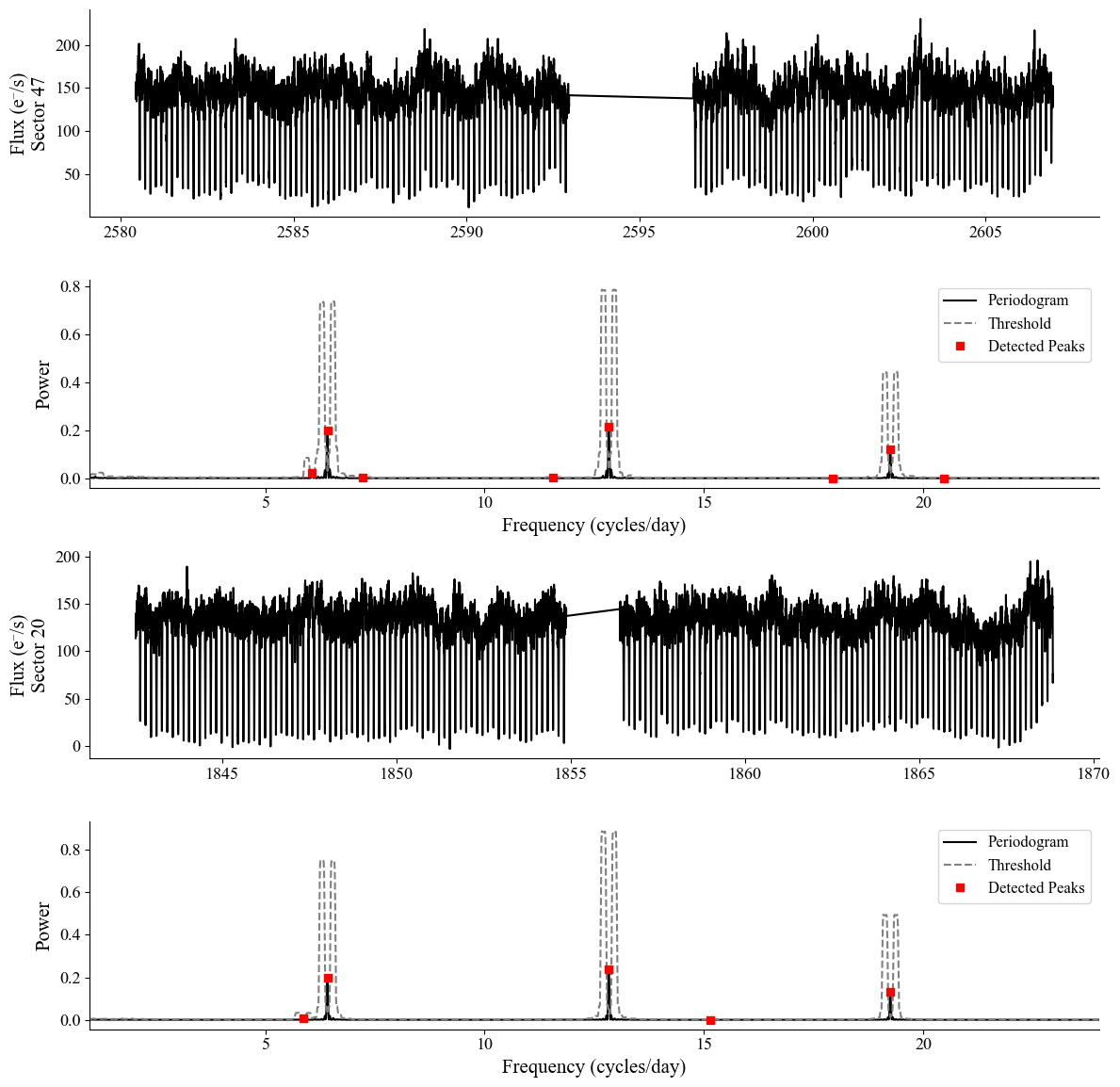}
    \caption{\textit{TESS} Sector 20 and Sector 47 light curves and periodograms of BH Lyncis (TIC ID  276252961), a nova-like cataclysmic variable known to exhibit both positive and negative superhumps. In Sector 20, a dominant signal is detected at approximately 5.86 cycle d$^{-1}$, and in Sector 47, at 6.06 cycle d$^{-1}$, both of which are lower than the known orbital frequency (6.41 cycle d$^{-1}$) and are consistent with positive superhumps. Additionally, a weaker but distinct signal at 6.62 cycle d$^{-1}$ is present, corresponding to a negative superhump.}
    \label{fig:BH_lyn}
\end{figure}

An example of this case is presented in Fig.\ref{fig:BH_lyn}. The figure shows the light curves and periodogram analyses of BH Lyncis (TIC ID  276252961) based on \textit{TESS} observations from Sector 20 and Sector 47. BH Lyn is an eclipsing nova-like cataclysmic variable (CV) with an orbital period of approximately 3.74 hours, corresponding to 6.415 cycles per day $^{-1}$. \citep{bruch2024_SUPERHUMP, Stanishev2007}. This frequency was successfully recovered by the algorithm.

However, since the system exhibits both positive and negative superhumps, the superhump signals interfere with the accurate identification of the first fundamental frequency and, in some cases, lead to incorrect assignment of the second fundamental. In fact, two different first fundamental frequencies were identified in the two sectors: in Sector 47, a positive superhump was detected at 6.06~cycle d$^{-1}$, whereas in Sector 20, it appeared at 5.86~cycle d$^{-1}$. For sources affected by such ambiguities, two approaches were adopted. If concerns regarding the signal–to–noise reliability were present, the sources were excluded from the final catalogue. When the detected signals were considered sufficiently robust, an explicit note indicating the possibility that the 1st fundamental of the system corresponds to a superhump was added in the Extra Info column (see Table \ref{tab:random10_cv}); this value was not removed from the 1st fundamental column, but it was also \textit{not} listed in the $f_{\rm orb}$ column. 

\subsection{CV Confident Catalogue}

The CCC comprises a total of 910 CVs and candidate sources. A sample output of the catalogue is provided in Table~\ref{tab:random10_cv}.
For each source, the catalogue lists the TESS Input Catalog Identifier (TIC ID), target name, and the \textit{TESS} observing sectors.
It reports the first fundamental frequency detected by the algorithm, the second fundamental frequency when present, and their associated uncertainties.
The presence of negative and positive superhumps is indicated as yes or no, followed by an Extra info field for special remarks and an Appendix flag identifying sources discussed in greater detail in the Appendix of this paper.
Also, a separate $f_{\rm orb}$ column provides the values of the first fundamental frequency that are considered to represent the orbital period. In some cases $f_{\rm orb}$ is different from the 1st fund column value because the latter is not always interpreted as the orbital period. In addition, Galactic coordinates have been included in the catalogue (RA and Dec).

The identification of superhumps was not performed by the algorithm, but rather relied entirely on visual inspection. Notably, the presence of superhumps occasionally led the algorithm to misidentify the true fundamental frequency. This became the primary reason why visual inspection was required across the entire catalogue.

If an additional frequency detected by the algorithm has been unequivocally established in the literature as the orbital period, and we consider this value to be reliable, the first fundamental frequency reported by the algorithm was replaced by this confirmed value.
Such a substitution was made only when the algorithm itself securely detected the literature value. The values in the  $f_{\rm orb}$ column are only those detected in the analysis, not values taken from the literature. In cases where a frequency confidently identified in the literature as the orbital period was neither the 1st fundamental nor detected by the algorithm, the source was either included in the appendix for further discussion or the $f_{\rm orb}$ entry was left blank.

\begin{figure}
    \centering
    \hspace*{-0.05\textwidth} 
    \includegraphics[width=0.5\textwidth]{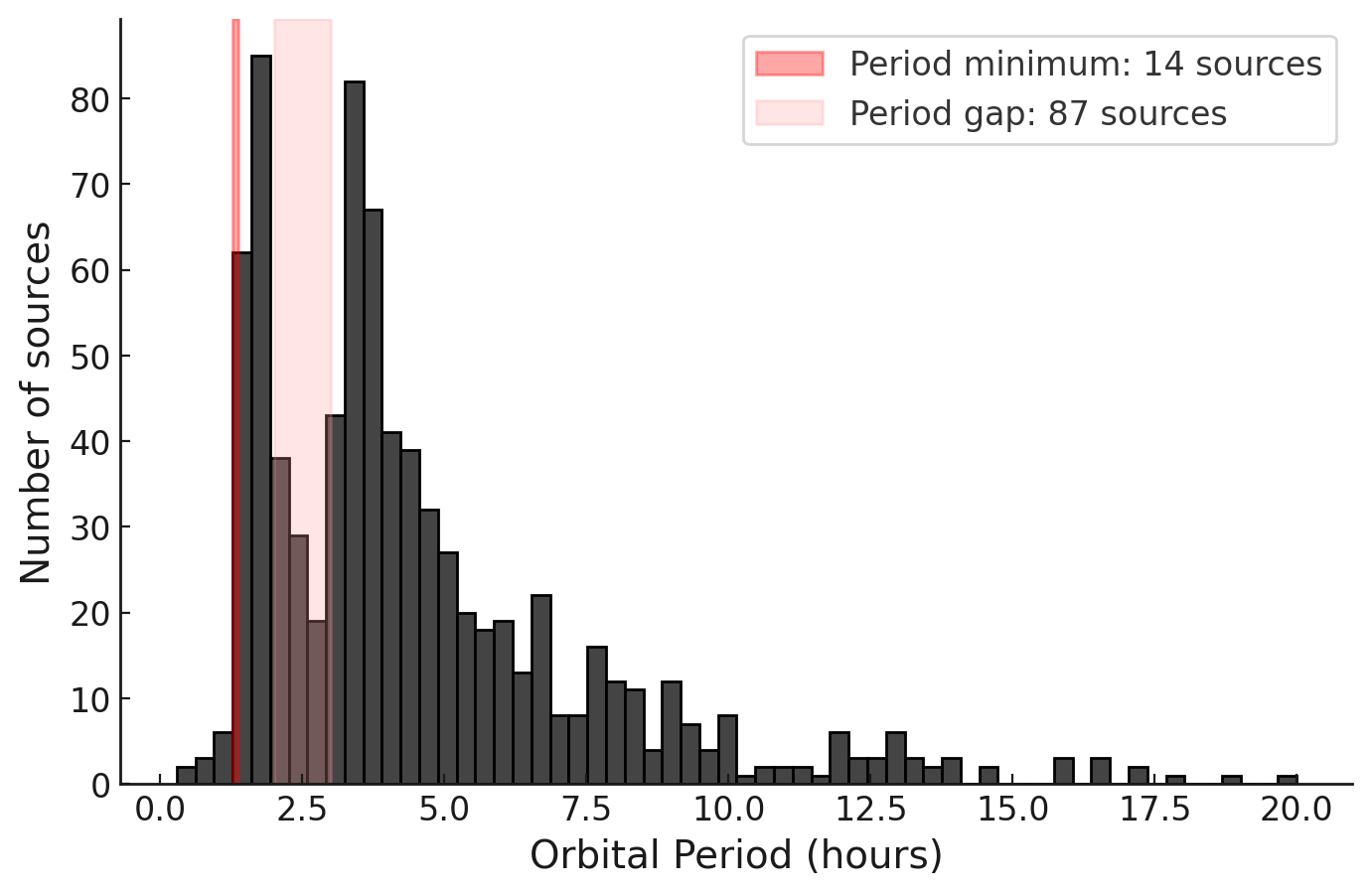}
    \caption{Histogram of the orbital period $P_{\rm orb}$ (computed from $f_{\rm orb}$ column), highlighting the Period Minimum and the Period Gap.}
    \label{fig:orb_dist}
\end{figure}

Figure~\ref{fig:orb_dist} shows the distribution of likely orbital periods for the 910 sources in the CCC. The distribution clearly reflects the evolutionary features of the CV population \citep{ritter2003catalogue}. A significant drop in the number of systems within the 2--3 hour range (pink region) supports the presence of the well-known period gap observed in CV evolution. In the CCC, 87 systems were found to lie within this gap. Additionally, an accumulation of systems around $\sim$1.33 hours (82 minutes) corresponds to the observational period minimum limit (red region). 
Systems located below the period minimum that are known or suspected to be AM CVn stars have been included in the Appendix. A detailed study of systems with orbital periods below 82 minutes will be presented in Mendoza et al. (in prep.).

In addition, we have identified 7 new magnetic systems and candidates, for which we provide precise determinations of orbital periods and, for the IPs, their corresponding spin periods (see the Appendix for a brief overview).

Additionally, systems for which a note is required such as cases where the algorithm detected a first or second harmonic instead of a true fundamental frequency, or systems exhibiting clear evidence of contamination are documented in the Appendix as well.

\subsection{Comparison with some recent TESS-based catalogues}

In recent years, a growing number of systematic studies have investigated cataclysmic variables using TESS light curves, leading to the publication of several subtype-specific catalogues. These works have provided statistically robust samples and have significantly improved our understanding of orbital, spin, and side-band periodicities across different classes of CVs, particularly magnetic systems. Within this context, it is essential to compare newly developed catalogues with recent \textit{TESS}-based studies in order to assess consistency and quantify overlaps.

\citet{Hernandez_2025} performed a systematic search for orbital periods of 95 magnetic CVs from the \citet{Schwope2025PolarCat} catalogue and with available \textit{TESS} two-minute cadence light curves. Their methodology combines four independent period-search techniques, namely Lomb–Scargle periodogram, autocorrelation function, sine fitting, and Fourier power spectrum analysis. The reliability of the detected periods was evaluated based on consistency across different methods and TESS sectors. In addition, \citet{Hernandez_2025} assessed the reliability of period detections by testing the recoverability of a periodic signal in simulations of \textit{TESS} light curves with varying noise levels, thereby establishing a probabilistic framework that quantifies the detection probability of a periodic signal as a function of its signal-to-noise ratio in the power spectral density of the light curve.

Cross-matching the 95-source sample presented by \citet{Hernandez_2025} with the CCC catalogue results in 81 common objects. For the majority of these overlapping systems, the derived orbital frequencies show good agreement between the two studies.

Discrepancies are found in two systems, Paloma (TIC 369210348) and IGR J19552+0044 (TIC 228975750), for which the dominant photometric modulation detected by the CCC pipeline does not correspond to the orbital period. In both cases, the first fundamental frequency identified by CCC is associated with non-orbital variability, consistent with the interpretations discussed by \citet{Hernandez_2025}. A detailed discussion of these systems is provided in Appendix.

In contrast, CCC shows direct agreement with \citet{Hernandez_2025} for one system for which the orbital period was reported outside the previously established literature. For J0733+2619 (TIC 94840820), the orbital frequency measured by CCC (7.170 cycle day$^{-1}$) is nearly identical to the value reported by \citet{Hernandez_2025} of 7.1717 cycle day$^{-1}$. 

\citet{Bruch2024} presents a study that is directly relevant to the present work and is based on the analysis of cataclysmic variables using TESS light curves. The study focuses on improving or revising the orbital periods of 53 CV systems through the combined use of TESS and Kepler photometry. The results demonstrate that high–time-resolution, space-based observations can refine previously reported orbital periods and, in some cases, lead to substantial revisions of values adopted in the literature. A significant fraction of the analysed systems exhibit updated or corrected orbital solutions.

We did not perform a direct cross-match between \citet{Bruch2024} dataset and the CCC catalogue, as the published tables are not available in a machine-readable format and would require manual compilation. However, in the Appendix, we explicitly identify the systems for which the orbital period values differ from those reported in earlier literature and were revised by \citet{Bruch2024}. In this sense, \citet{Bruch2024} study serves as an independent validation of our results.

In particular, for the four systems that show mismatches with the Ritter \& Kolb catalogue, we find orbital period values consistent with those reported by \citet{Bruch2024}. In addition, we expect that many other sources for which we report orbital periods differing from the literature, and which we considered not to require individual appendix discussion, are also included in \citet{Bruch2024} sample. However, due to the size of the dataset and the absence of machine-readable tables, it is not straightforward to provide a reliable quantitative estimate of the total overlap.

\subsection{Cross-Match with the Ritter \& Kolb Catalog}

The Ritter \& Kolb catalogue \citep{ritter2003catalogue} is one of the most widely used reference databases for cataclysmic variables and related interacting compact binaries. It provides a comprehensive compilation of systems reported in the literature, including orbital periods, system classifications, photometric properties, and bibliographic references. The catalogue incorporates measurements obtained from a wide range of observational techniques, such as time-resolved photometry, spectroscopy, and radial-velocity studies, collected over several decades.

The Ritter \& Kolb catalogue is not based on a homogeneous re-analysis of archival data. Instead, it represents a curated literature summary in which orbital period values are adopted from individual published studies, often derived using different instruments, observing strategies, and analysis methods. As a result, the reported periods may vary in precision and reliability, particularly for systems with sparse observational coverage, strong aliasing, or complex variability behaviour.

To validate the CV Confident Catalogue against existing literature, we performed a cross-match with the most recently updated version of the Ritter \& Kolb catalog \citep{ritter2003catalogue}. This cross-matching was based on RA and DEC coordinates, resulting in 300 sources having positional matches. Among them, 215 sources exhibit a one-to-one match when considering a frequency uncertainty of 0.05 cycle d$^{-1}$. The frequency tolerance of $0.05~\mathrm{cycle~d^{-1}}$ was chosen to
reflect the intrinsic frequency resolution of a typical TESS sector ($\Delta f \sim 1/T \approx 0.04~\mathrm{cycle~d^{-1}}$ for $T \approx
27$~days), while also accounting for the heterogeneous uncertainties in orbital periods reported in the Ritter \& Kolb catalogue.

However, 85 sources exhibit discrepancies between the two catalogues. In these cases, strong periodic signals are found to be inconsistent with the values reported in the R\&K catalogue, suggesting that those catalogue entries may need to be revised. For each of these sources, a literature review was conducted to determine whether any corrections had been proposed previously. Both previously suggested revisions and new corrections proposed for the first time in this study, covering a total of 80 sources, are presented in the Appendix as R\&K mismatches. For 39 sources, the values reported in the R\&K catalogue were updated, with the revised periods justified by independent measurements and corroborating studies in the literature.

\section{Conclusion}
\label{sec:conclusion}

In this study, \textit{TESS} light curves of 2544 cataclysmic variables (CVs) and CV candidates, proposed through the \textit{TESS} Guest Investigator program, were analysed using an automated algorithm to generate periodograms and identify periodic signals. After an additional visual inspection, 910 sources were selected and compiled into a catalogue. For each of these, the frequency believed to represent the orbital period (the first fundamental), a possible spin frequency (if present, as a second fundamental), and, where applicable, visually confirmed superhump features were included. The catalogue is available online via VizieR.

The appendix also lists systems whose previously reported orbital periods appear to be incorrect and should be updated, based on this work and other supporting studies. The goal of this catalogue is to provide more accurate period measurements of CV systems using \textit{TESS} data, to introduce potential new CV candidates to the literature, and to contribute to studies on the evolution of these systems. 

This catalogue has been designed as a living resource. The current release covers \textit{TESS} observations up to Cycle 6 (ended on 1 October 2024), incorporating all available sectors within this cycle. Future updates will include data from subsequent \textit{TESS} cycles, enabling both the addition of newly observed targets and the refinement of parameters for previously listed sources. Moreover, more than a thousand objects from the initial list remain to be analysed, and as noted earlier, the properties of the sources removed from the current sample do not necessarily indicate that the detected signals were spurious. We therefore plan to revisit these sources in future updates.

\section*{Acknowledgements}
M.K.D gratefully acknowledges support from the \textit{Durham Physics Doctoral Studentship} awarded by Durham University. S.S. acknowledges support by STFC grant ST/T000244/1 and ST/X001075/1. Z.A.I gratefully acknowledges support from the UK Research and Innovation's Science and Technology Facilities Council (STFC) grant ST/X508767/1. S.H.D acknowledges financial support from Deutsche Forschungsgemeinschaft (DFG) under grant number STE\,1068/6-2. K.I. was supported by the Polish National Science Centre (NCN) grant 2024/55/D/ST9/01713. Y.C. acknowledges support from the grant RYC2021-032718-I, financed
by MCIN/AEI/10.13039/501100011033 and the European Union NextGenerationEU/PRTR. L.E.R.S, W.M. and J.K. acknowledge support from NASA grants NNH22ZDA001N-6152 and 80NSSC24K0638. M.P.M is partially supported by the Swiss National Science Foundation
IZSTZ0$\_$216537 and by UNAM PAPIIT-IG101224.

\section*{Data Availability}
The \textit{TESS} data used in the analysis of this work is available on the MAST webpage \url{https://mast.stsci.edu/portal/Mashup/Clients/Mast/Portal.html}.



\bibliographystyle{mnras}
\bibliography{main} 




\appendix

\section*{\Large\bfseries Appendix}
\label{sec:appendix}

\begin{table*}
\centering
\caption{The 11 excluded sources and their multi-sector frequency detections, despite consistent signals being identified across multiple TESS sectors.}
\label{tab:multi_sector_19}
\begin{tabular}{l l c c}
\hline
TIC ID & Name & Frequency (c/d) & Sectors \\
\hline
230861869  & V* V1329 Cyg            & 2.05   & 55;75 \\
233738348  & RX J1831.7+6511         & 5.97   & 78;80 \\
252951144  & GALEX J070859.5+520309  & 7.79   & 20;47;60 \\
275134211  & ASASSN-17od             & 9.47   & 46;72 \\
306023513  & V*LZ Mus                & 6.11   & 37;38   \\
312474753  & Gaia DR2 345111707114997312 & 4.01   & 43;45   \\
352302825  & V* V1006 Cyg            & 20.18  & 41;56;57;77  \\
412059865  & CRTS J051922.9+155435   & 5.48   & 43;45;71   \\
422947065  & V* FS Cet               & 1.43   & 31;42;43;71  \\
764302436  & V* DM Gem               & 8.07   & 43;44   \\
1807414830 & Gaia DR2 4514101337822714240  & 1.35  & 40;53   \\
\hline
\end{tabular}
\end{table*}

\subsection*{Mismatches with Ritter and Kolb}

\subsubsection*{\textbf{2MASS J08384911-4831248 
(TIC 93196973)}}
The R\&K quotes an orbital period of 7.92 hr (3.03 cycle d$^{-1}$). Upon closer inspection, this period has been retrieved from \citet{bernardini2012}, where the orbital period is quoted as $8.0\pm1.0$ hr. This measurement has been inferred from the optical sideband to the X-ray detected spin period of $1560\pm7$ s. Our analysis reveals strong signals at 6.10 hr (3.936 cycle d$^{-1}$)  which we interpret as the orbital period and 1598.5s which we interpret as the spin modulation. The \textit{TESS} periodograms also reveal strong signals at the expected sidebands. The X-Rays support the optical agreement with the interpretation of IGR J08390$-$4833 as a magnetic CV, specifically an intermediate polar. The different orbital and spin signals are interesting and may suggest that the WD in IGR J08390$-$4833 is spinning down and, if so, at a rate of $\approx 1.5 \times 10^{-7}$s s$^{-1}$. This value is very high for IPs, and we also note that throughout all 6 \textit{TESS} sectors over 5 years, the 1598.5 s modulation has remained unchanged.

\subsubsection*{\textbf{V* V2275 Cyg (TIC 289786644)}}

The R\&K quotes 7.55 hr (3.178 cycle d$^{-1}$) based on \citet{Solen_v2275} who detected the signal using 8 nights of optical photometry. Their periodogram shows a prominent signal with strong daily aliasing. In addition, \citet{Solen_v2275} (see also Balman et al.) reported a second periodicity at $\sim$58.6 cycle d$^{-1}$ (24.6 min). In our \textit{TESS} data, we observe a peak near this frequency; however, its power is comparable to the surrounding noise level, and peaks of similar strength are present at neighboring periods. We therefore do not consider the $\sim$58.6 cycle d$^{-1}$ signal to be significant in our data. We find a period of 11.11 hr (2.169 cycle d$^{-1}$) and the related first harmonic, which we interpret as the true orbital period of this system, closely matching the lower frequency signal reported by \citet{Solen_v2275}.

\subsubsection*{\textbf{U Leo (TIC 70680826)}}
The R\&K quotes 6.418 hr (3.739 cycle d$^{-1}$) based on \citet{downes89}. They report sinusoidal photometric variability in the R band, suggesting two possible orbital periods: 3.21 hr or 6.42 hr. Our \textit{TESS} analysis reveals a strong signal at 3.204 hr (7.49 cycle d$^{-1}$), which we interpret as the first harmonic of the quoted 6.42 orbital period found in \citet{downes89}.

\subsubsection*{\textbf{2MASS J05181432+2941130 (TIC 62257830)}}
The R\&K catalog quotes an orbital period of 5.719 h (4.19 cycle d$^{-1}$) based on \citet{Witham2007}, who detected the period through radial velocity. Their periodogram is heavily affected by daily aliasing. Our \textit{TESS} analysis reveals a strong signal in this eclipsing system at 4.95 hr (4.8529 cycle d$^{-1}$) and its related 11 harmonics. We interpret this value as the orbital period.

\subsubsection*{\textbf{V344 Ori (TIC 434389111)}}
The R\&K catalog quotes an orbital period of 5.62 hr (4.277 cycle d$^{-1}$) based on radial velocity measurements of \citet{throrstensen10}. We find a strong signal at 2.76 hr (8.709 cycle d$^{-1}$) in \textit{TESS} which may be interpreted as the harmonic of this reported orbital period. However, we note that this signal is not exactly twice that reported in \citet{throrstensen10}, leaving the true orbital period for this system somewhat ambiguous. Moreover, \citet{thorstensen10} reported a range of radial-velocity aliases rather than a uniquely determined value. 
One possibility is that \textit{TESS} detects a harmonic signal of the superhump modulation in this outbursting dwarf nova system, and the reported  5.62 hr of \citet{throrstensen10} is the true orbital period.

\subsubsection*{\textbf{NR TrA (TIC 1211418907)}}
The R\&K catalog quotes an orbital period of 5.26 h (4.19 cycle d$^{-1}$)  based on the analysis of its light curve derived through photometric observations \citep{walter2012}. We cannot detect or see this period in either of the two observed TESS sectors.
Our \textit{TESS} analysis reveals a strong signal in 13.19 hr (1.829 cycle d$^{-1}$), and related  harmonics as also found in \citet{Bruch2024}. Given the \textit{TESS} sampling clearly showing eclipses in this system, we interpret the 13.19 hr signal as the orbital period.

\subsubsection*{\textbf{V825 Her (TIC 101675240)}}
The R\&K catalog quotes an orbital period of 4.94 h (4.85 cycle d$^{-1}$) based on the radial velocity study of \citet{Ringwald2005}. We find a strong signal in 6.12 hr (3.9208 cycle d$^{-1}$), which we interpret as the orbital period, also consistent with \citet{Bruch2024}.

\subsubsection*{\textbf{V1323 Her (TIC 75654576)}}
The R\&K quotes an orbital period of 4.40 hr (5.45 cycle d$^{-1}$) for V1323 Her based on the radial velocity study by\citet{anzolin2008}. Their periodogram is heavily affected by daily aliasing. We find a signal at 4.85 hr, (4.953 cycle d$^{-1}$), which is within the uncertainty range of the 1/2-day alias mentioned by \citet{anzolin2008}, and interpret this signal as the orbital period, also consistent with \citet{Bruch2024}.

\subsubsection*{\textbf{Mis V1448 (TIC 233264291)}}
The R\&K quotes an orbital period of 4.40 hr (5.45 cycle d$^{-1}$) for Mis V1448 Her based on photometry and reported in a VSNET-alert \citep{denisenko2012vsnet_MIS}. Our \textit{TESS} analysis reveals a signal at 5.16 hr (4.661 cycle d$^{-1}$) which we interpret as the orbital period.

\subsubsection*{\textbf{FO Per (TIC 267052599)}}
The R\&K quotes an orbital period of 4.13 hr  from a radial velocity study of \citet{sheets07} which is highly affected by daily aliasing. Our \textit{TESS} analysis reveals a signal at 3.39 hr (7.107 cycle d$^{-1}$) and related harmonics, which we interpret as the orbital period.

\subsubsection*{\textbf{2MASS J09275195-3910524  (TIC 4970499)}}
The R\&K quotes an orbital period of 4.1 hr (5.853 cycle d$^{-1}$) based on the radial velocity study of \citet{pretorius2008_hALPHA}. Their spectroscopy was observed only for one night and the signal in the periodogram is thus very broad. We detected a signal at 5.42 hr (4.4381 cycle d$^{-1}$)  and related harmonics, as well as a signal and its related harmonics at 6.44 c/d, consistent with that of \citet{pretorius2008_hALPHA}. The signal at 4.4381 cycle d$^{-1}$ signal exhibits the first through seventh harmonics, with the harmonic components often comparable in power to the fundamental, whereas the 6.44 cycle d$^{-1}$ signal shows only a weak first harmonic relative to the fundamental. Moreover, the 4.4381 cycle d$^{-1}$ signal itself is stronger than the 6.44 c/d signal. It is unclear what these correspond to. Further observations and analysis will be required to confirm the nature of this signal.


\subsubsection*{\textbf{V* V842 Cen (TIC 411603422)}}
The R\&K catalogue quotes an orbital period of 3.94 hr (6.0914 cycle d$^{-1}$) for V842 Cen based on the value reported by \citet{woudt2009} based on optical photometry heavily affected by daily aliasing. We find a strong signal at 3.55 hr (6.751 cycle d$^{-1}$), which aligns with the range reported by \citet{Epic_OM_v842cen} in their study of V842 Cen. It is important to note that this system is a nova-like, most likely face-on, and it is unclear whether the observed signal in \textit{TESS} data is the orbital period or a negative superhump.

\subsubsection*{\textbf{V* CM Del (TIC 305247484)}}
The R\&K quotes an orbital period of 3.89 hr (6.17 cycle d$^{-1}$) based on the value reported by \citet{shafter1985_CM_dEL}. They also noted that the phase coverage was limited and the velocity amplitude was poorly constrained, making the period estimate uncertain.
We found a signal at 3.38 hr (7.0902 cycle d$^{-1}$), as well as related harmonics for this eclipsing system. We suspect this is the true orbital period of the binary, as also suggested by \citet{Bruch2024}.

\subsubsection*{\textbf{J2337+4308 (2MASS J23375921+4308509) (TIC 176794745)}}

The R\&K catalogue lists an orbital period of 3.852 hr (6.230 cycle d$^{-1}$) based on \citet{Thorstensen2002}. A later spectroscopic study by \citet{Weil2018} reported a period of 0.130 d (3.13 hr, 7.67 cycle d$^{-1}$) for this system. The object is classified as a VY Scl type cataclysmic variable and is known to exhibit pronounced high and low brightness states \citep{dUFFY}. Our analysis of the TESS data reveals a coherent modulation at 11.94 hr (2.0137 cycle d$^{-1}$), accompanied by four harmonics. Given the presence of multiple spectroscopic period measurements and the characteristic behaviour of VY Scl stars, the 11.94 hr signal cannot be interpreted as a harmonic of any of the previously reported orbital periods. Additionally, a nearby source within 20 arcseconds, whose magnitude and variability type are unknown, introduces a risk of photometric contamination. For this reason, we flag this system as potentially contaminated in our catalogue.

\subsubsection*{\textbf{ATO J119.2035-12.7816 (TIC 56380389)}}
The R\&K quotes an orbital period of 3.49 hr (6.88 cycle d$^{-1}$) based on the value reported by \citet{woudt2010} where the authors interpret their result as tentative. We find a strong signal at 10.71 hr (2.236 cycle d$^{-1}$), as well as the related first harmonic and interpret this as the orbital period.

\subsubsection*{\textbf{V393 Hya (TIC 382378369)}}
The R\&K quotes 3.23 hr based on optical photometry from the Center for Backyard Astrophysics. We find a strong signal at 2.95 hr (8.141 cycle d$^{-1}$) in 2 \textit{TESS} sectors, while we find a negative superhump signal at 7.9 cycle d$^{-1}$, in sector 63. We interpret the 2.95 hr signal as the orbital period.

\subsubsection*{\textbf{2MASS J01303186+6221324 (TIC 389536783)}}
The R\&K quotes 3.12 hr based on radial velocity measurements highly affected by daily aliasing of \citet{Witham2007}. We find a strong signal at 3.58 hr (6.6959 cycle d$^{-1}$) in 2 \textit{TESS} sectors, while we find a negative superhump signal at 7.02 cycle d$^{-1}$, and related superorbital modulation at 0.33 cycle d$^{-1}$, in sector 58. Unambiguously confirming the orbital period of 3.58 hr.

\subsubsection*{\textbf{BX Pup (TIC 128898689)}}
The R\&K quotes 3.05 hr (7.868 cycle d$^{-1}$) based on \citet{Augusteijn1993priv}. We find a strong signal at 5.31 hr (4.518 cycle d$^{-1}$) in 2 \textit{TESS} sectors and the first harmonic. However, an outburst is observed in both of the observed sectors. Therefore, we cannot state with certainty that our signal corresponds to the orbital period.


\subsubsection*{\textbf{LQ Pegasi (TIC 305267248)}}
The R\&K quotes 2.99 hr (8.03 cycle d$^{-1}$) based on optical photometry from \citet{papadaki2004} affected by daily aliasing. We find a strong signal at 3.44 hr (6.9813 cycle d$^{-1}$) and the related harmonics and interpret this as the orbital period.

\subsubsection*{\textbf{V592 Cas (TIC 403308597)}}
The R\&K catalogue quotes 2.76 hr (8.70 cycle d$^{-1}$) based on optical photometry from \citet{taylor98}. We find a strong signals at 2.94 hr (8.17 cycle d$^{-1}$) and at 2.76 hr (8.68 cycle d$^{-1}$), and thus confirm the orbital period for this system as found in \citet{taylor98}.

\subsubsection*{\textbf{2MASS J19241081+4459348 (KIC 8751494) (TIC 159580213)}}

The R\&K catalog quotes 2.75 hr (8.74 cycle d$^{-1}$) based on optical photometry obtained with \textit{Kepler} by \citet{Kato_2013_KIC87514994}, who also reported negative superhumps in the range of 9.25–9.34 cycle d$^{-1}$. We find a strong signal at 8.62 hr (2.78 cycle d$^{-1}$) and related harmonics, where the folded \textit{TESS} light curves appear to show (partial) eclipses.

However, \citet{Bruch2023} reports that the \textit{TESS} light curves of this source are strongly contaminated by the nearby variable ATO J291.0335+44.9915, located only $\sim$42 arcsec away (approximately twice the \textit{TESS} pixel scale). The power spectra of this neighbouring object dominate the \textit{TESS} data with a fundamental frequency at 5.56 d$^{-1}$ and its harmonics. Consequently, part of the variability detected in the \textit{TESS} observations, especially the eclipsing-like modulation, likely originates from this contaminating source. Therefore, we classify TIC 159580213 as a potentially contaminated system in our analysis.

\subsubsection*{\textbf{2MASS J07381772+2855198 (TIC 95074141)}}
The R\&K catalogue quotes 2.10 hr (11.43 cycle d$^{-1}$) based on radial velocity measurements of \citet{szkody03} using 8 spectra. We identify a strong signal at 5.55 hr (4.32 cycle d$^{-1}$) and related harmonics for this (partial) eclipsing system, which we identify as the orbital period.

\subsubsection*{\textbf{RE J1002-19 (TIC 875790341)}}
The R\&K catalogue quotes 1.78 hr (13.46 cycle d$^{-1}$) based on ROSAT observations of \citet{beuermann95}, where the authors explicitly state the orbital period determination is preliminary. We identify a strong signal at 1.67 hr (14.414 cycle d$^{-1}$) and the related first harmonic, which we interpret as the orbital period of this system.

\subsubsection*{\textbf{V823 Cyg (TIC 40223843)}}
The R\&K catalogue quotes 1.06 hr (22.64 cycle d$^{-1}$) based on \citet{Voloshina2014vsnet17970} where possible positive superhumps were detected. We identify a strong signal at 4.53 hr (5.312 cycle d$^{-1}$) and related first harmonic. Additionally, a source, ZTF J200149.20+360729.2, lies 5 arcsec away and is reported to have an orbital frequency of $\sim2.6$ cycle d$^{-1}$ (citation to be added). Our peak at 5.312 cycle d$^{-1}$ is consistent with the second harmonic of that frequency; thus, so contamination cannot be excluded, and the true orbital period of the target remains uncertain.

\subsubsection*{\textbf{ASASSN 15ea (TIC 390207847)}}
The R\&K catalogue quotes 1.97 hr (12.20 cycle d$^{-1}$) based on \citet{vanmunster2015_vsnet18357_Assasn_15ea}. where possible positive superhumps were detected. We identify a strong signal at 12.29 hr (1.947 cycle d$^{-1}$) and related harmonics in several \textit{TESS} sectors which we interpret as the orbital period of this system, as also confirmed in \citet{refId0_ASASSN15ea}.

\subsubsection*{\textbf{BK Lyn (TIC 8765832)}}
The R\&K catalogue quotes 1.80 hr (13.34 cycle d$^{-1}$) based on \citet{ringwald96} radial velocity measurements. We identify a strong signal at 12.70 cycle d$^{-1}$, and a weaker one at 13.343 cycle d$^{-1}$, the same frequency of \citet{ringwald96}. We interpret the 12.70 cycle d$^{-1}$, as the negative superhump of this and the 13.34 cycle d$^{-1}$ as the orbital period.

\subsubsection*{\textbf{RX J0150.8+3326 (TIC 620725990)}}
The R\&K catalogue quotes 1.70 hr (14.08 cycle d$^{-1}$) based on
\citet{Kato2012vsnet15026}. We identify a strong signal at 5.00 hr (4.804 cycle d$^{-1}$) in two \textit{TESS} sectors which we interpret as the orbital period of this system.

\subsubsection*{\textbf{CRTS J001538.3+263657 (TIC 306066903)}}

The R\&K catalogue quotes 1.68 hr (14.29 cycle d$^{-1}$) based on a radial velocity study of \citet{szkody14}. We identify a strong signal at 2.74 hr (8.775 cycle d$^{-1}$), but this appears to be blended within a broad-band noise component, reminiscent of the quasi-periodic oscillations reported by \citet{veresvarska24} for other systems.
We cannot confidently interpret the 2.74 hr modulation as the orbital period of the system.


\subsubsection*{\textbf{CRTS J214426.4+222024 (TIC 330158146)}}
The R\&K catalogue quotes 1.54 hr (15.63 cycle d$^{-1}$) based on the VSNET-alert of Kato. T. Our \textit{TESS} analysis reveals a signal at 3.68 hr (6.518 cycle d$^{-1}$) and related first harmonic with higher power than the fundamental, which we interpret as the orbital period for this system.

\subsubsection*{\textbf{V476 Peg (TIC 399872907)}}
The R\&K catalogue quotes 1.53 hr (15.70 cycle d$^{-1}$) based on photometric observations of \citet{antipin04}. We find a strong signal at 4.41 hr (5.46 cycle d$^{-1}$) and its first harmonic, which we interpret as the orbital period for this dwarf nova outbursting system.

\subsubsection*{\textbf{FS Aur (TIC 76381942)}}
The R\&K catalogue quotes 1.43 hr (16.78 cycle d$^{-1}$) based on optical photometry of \citet{neustoev12}. We find a strong signal at 3.43 hr (7.008 cycle d$^{-1}$) and related first harmonic. This signal has also been reported by \citet{neustoev12} and interpreted as a ``long photometric period''. We do not observe any signal at the claimed orbital period of FS Aur, and cannot decisively infer the orbital period for this system.

\subsubsection*{\textbf{ASASSN 14cl  (TIC 2001466142)}}
The R\&K catalogue quotes 1.40 hr (17.13 cycle d$^{-1}$) based on optical photometry from \citet{Oksanen2014vsnet17398}. We find a strong signal at 13.80 hr (1.737 cycle d$^{-1}$) and related harmonics for this partially eclipsing system, including a harmonic at 17.43 cycle d$^{-1}$, potentially coinciding with the signal reported by \citet{Oksanen2014vsnet17398}.

\subsubsection*{\textbf{CRTS J173516.9+154708 (TIC 267353881)}}
The R\&K catalogue quotes 1.40 hr (17.15 cycle d$^{-1}$) based on optical photometry from \citet{Masi2011vsnet13470}. We find a strong signal at 4.24 hr (5.67 cycle d$^{-1}$) which we interpret as the orbital period of this system.

\subsubsection*{\textbf{IGR J19552+0044 (TIC 228975750)}}
The R\&K catalogue quotes 1.39 hr (17.23 cycle d$^{-1}$) based on a tentative period identification using radial velocity measurements of \citet{thorstensen13}. In a later paper \citet{tovmassian17} identified this target as an asynchronous polar with an orbital period of 1.39 hr and a 1.36 hr WD spin period.

Later in 2025, \citet{Hernandez_2025} reported a dominant frequency at 17.714 cycle d$^{-1}$, interpreted as the white dwarf spin period, and tentatively identify a weaker signal at 17.346 cycle d$^{-1}$ as the orbital frequency of the system. 
Our analysis picked 17.714 cycle d$^{-1}$, which is recorded as the first fundamental frequency. However, this frequency corresponds to the spin modulation rather than the orbital period. The orbital frequency reported in the literature is not recovered by our algorithm, likely due to its low amplitude relative to the dominant spin variability. We therefore retain the detected frequency as the first fundamental in our catalogue.

\subsubsection*{\textbf{RX J0154.0-5947 (TIC 231044546)}}
The R\&K catalogue quotes 1.33 hr (17.99 cycle d$^{-1}$) based on a poster presented by Burwitz, V., but a later work by \citet{beuermann21} reports an orbital period of 1.48 hr (16.189 cycle d$^{-1}$) using radial velocity measurements. Our \textit{TESS} analysis reveals the orbit as a strong signal and related harmonic at 16.19 cycle d$^{-1}$. We also identify a strong signal at 1.86 cycle d$^{-1}$, which may indicate a beat with an asynchronous WD spin period, tentatively making this system an asynchronous polar. However, the source lies only 8 arcsec from UCAC4 152-001731, and the possibility of contamination from this nearby star should therefore be taken into consideration.

\subsubsection*{\textbf{EQ Lyn (TIC 741679184)}}
The R\&K catalogue quotes 1.33 hr (18.12 cycle d$^{-1}$) based on 17 radial velocity measurements of \citet{mukadam13}. The authors also find a signal at 85.77 min (16.78 cycle d$^{-1}$) based on optical photometry. Our \tess\ analysis reveals one strong signal at 5.01 hr (4.793 cycle d$^{-1}$) with no power at the reported orbital or superhump signals. Given the closeby distance of EQ Lyn (310 pc) and its position close to the WD track in the Gaia colour-magnitude diagram, we expect this system to have a small orbital period and low mass transfer rate, as also revealed by the average spectra obtained by \citet{mukadam13}. As such, we think the orbital period determination of \citet{mukadam13} is correct, but we are uncertain on the origin of the long \tess\ signal. Potential interpretations may include the precession frequency of the disk, but it is unclear whether this can be attributed to a tilted retrograde precessing disk as the superhumps from \citet{mukadam13} seem to suggest an elliptical prograde precessing disk. Further observations to unravel the periodic signals in this source are recommended.

\subsubsection*{\textbf{SDSS J190817.07+394036.4 (KIC 004547333, TIC 121107327)}}
The R\&K catalogue quotes 18.08 min (79.62 cycle d$^{-1}$) based on \textit{Kepler} photometry reported in \citet{fontaine11} who classify this system as an AM CVn-type binary given its small orbital period. A further study by \citet{kupfer15} confirms the orbit of this system through radial velocity studies. We recover a strong signal and related first harmonic in 6 \tess\ sectors at 13.88 hr (1.733 cycle d$^{-1}$). Although the folded lightcurve on this coherent period shows a double-peaked profile reminiscent of orbital period modulations, the radial velocity measurements of \citet{kupfer15} rule out the \tess\ as being related to the orbit. Potential interpretation for this low frequency signal may be related to a superorbital disk precession frequency (either nodal or apsidal), but we note that the frequencies of these signals generally vary over time. In addition, there is a nearby source within 20 arcsec whose magnitude and variability type are not well constrained, so the possibility of contamination from this object should be considered. Further observations to reveal the nature of this low frequency signal are encouraged.


\subsubsection*{\textbf{V587 Lyr (TIC 121992913)}}
The R\&K catalogue quotes 6.58 hr (3.65 cycle d$^{-1}$) based on the radial velocity study of \citet{thorstensen10}. We recover a strong signal and related first harmonic at 6.54 hr (3.668 cycle d$^{-1}$), about 2 minutes faster than that reported by \citet{thorstensen10}. As the system has been observed by \tess\ during both quiescence and outburst, we are confident this signal is the orbital period, which is consistent with the error quoted by \citet{thorstensen10}.

\subsubsection*{\textbf{MASTER OT J162323.48+782603 (TIC 159500861)}}

The R\&K catalogue quotes 2.113 hr (11.36 cycle d$^{-1}$) estimated from superhump maxima using the O–C diagram method in \citet{kato2015_SU_UMA}.
We see a strong signal at 2.039 hr (11.778 cycle d$^{-1}$). The signal is consistently detected across multiple TESS sectors, along with its first harmonic, suggesting a stable and coherent modulation. Another signal at 11.28 cycle d$^{-1}$, appears only in Sector 59, which shows an outburst, and also exhibits a harmonic component. Based on this consistency  among sectors and the harmonic structure, we believe the 11.78 cycle d$^{-1}$ signal is the true fundamental frequency. 

\subsubsection*{\textbf{V344 Lyrae (TIC 123234016)}}

The R\&K catalogue quotes 2.11 hr (11.38 cycle d$^{-1}$) based on the value reported by \citet{Osaki2013}. They determined the orbital period from high-cadence photometry by applying Fourier analysis to the continuous light curve, combined with prewhitening to remove superhump signals and phase-folding to isolate the orbital modulation. Their analysis yielded a negative superhump frequency around 11.62 cycle d$^{-1}$. 
We detected a strong signal at 2.06 hr (11.644 cycle d$^{-1}$). This value is slightly offset from the reported negative superhump frequency, which could be due to differences in the data span, sector coverage, or variations in the superhump period over time. The R\&K quotes the correct orbital period of the system and our algorithm picked the negative superhump signal. 

\subsubsection*{\textbf{MASTER OT J055845.55+391533.4 (TIC 724028977)}}

The R\&K quotes 1.34 hr (17.85 cycle d$^{-1}$) based on the value reported by \citet{2023arXiv230407695K}, although the original link is no longer accessible. For this reason, the method used to determine the period is unknown. 
We detected a signal at 1.32 hr (18.179 cycle d$^{-1}$), but this detection is based on only a single sector (Sector 73). Furthermore, this sector contains significant data gaps and an outburst, which makes the reliability of this detection uncertain. Moreover, three neighbouring sources are located within 20 arcsec of the target, including an eclipsing binary (Gaia DR3 3458275643465151744). Such a crowded environment casts doubt on the reliability of the detected signal and suggests that contamination is likely.

\subsubsection*{\textbf{V503 Cyg (TIC 193895999)}} 

The R\&K quotes 1.87 hr (12.85 cycle d$^{-1}$) based on the value reported by \citet{Kato_2013}. The orbital period was determined by using time-resolved photometry and phase dispersion minimization of the light curve. 
We detected a strong, stable signal at 1.825 hr (13.182 cycle d$^{-1}$) persisting over nine sectors, which was recorded by the algorithm as an orbital period candidate. However, based on the value reported in \citet{Pavlenko2024_V503Cyg}, this signal is considered to be a negative superhump.




\subsubsection*{\textbf{V342 Cam (TIC 103439731)}}

The R\&K quotes 1.81 hr (13.27 cycle d$^{-1}$) based on the value reported by \citet{Shears2011_HS0417}. They determined the orbital period by analysing both orbital humps and stable superhumps during the 2008 superoutburst, using time-series CCD photometry and O–C analysis. 

\textit{TESS} data reveals a strong signal at 1.82 hr (13.216 cycle d$^{-1}$) and its first harmonic among nine sectors. Only during Sector 25 we find an additional strong signal at 12.84 cycle d$^{-1}$. We interpret the 13.216 cycle d$^{-1}$ as the orbital period of the system and the 12.84 cycle d$^{-1}$ as a positive superhump.

\subsubsection*{\textbf{V* SS UMi (TIC 272781244)}}

The R\&K quotes 1.626 hr (14.75 cycle d$^{-1}$) based on the value reported by \citet{Kato_2013}. Orbital period was determined from light curve analysis of the 2012 observations. 

We detected a signal at 1.68 hr (14.272 cycle d$^{-1}$). But according to \citet{Kato_2013}, a superhump period during the 2004 superoutburst was measured as 1.68 hr (14.25 cycle d$^{-1}$) which is close to the signal our algorithm detected. We believe that the value in the R\&K catalogue is correct and that the value from our algorithm is incorrect, as the signal we detected is not consistently observed across the sectors.

\subsubsection*{\textbf{V* SX LMi (TIC 165690666)}}

The R\&K quotes 1.61 hr (14.88 cycle d$^{-1}$) based on the value reported by \citet{wagner1998sx}. Orbital period was determined from H$\alpha$ radial velocity variations in quiescence using the double-Gaussian convolution technique.  

We detected a signal at 1.67 hr (14.39 cycle d$^{-1}$) in only one sector and during an outburst. The same signal has also been reported in \citet{kato2009superhumps} as a positive superhump signal. Since the signal is not consistently observed across the available sectors and an outburst occurred in the relevant sector, we consider the finding of \citet{kato2009superhumps} to be correct and regard this signal as a positive superhump.

\subsubsection*{\textbf{V* V1240 Her (TIC 376790118)}}

The R\&K quotes 1.6 hr (14.92 cycle d$^{-1}$) based on the value reported by \citet{szkody2006cataclysmic}. The orbital period was determined from time-resolved photometry by measuring the sinusoidal modulation in the light curve. 
We detected a signal at 1.65 hr (14.577 cycle d$^{-1}$) in only a single sector.Therefore, we do not consider it reliable. Nevertheless, we decided to include it in the table.

\subsubsection*{\textbf{V* IX Dra (TIC 236763903)}}

The R\&K quotes 1.59 hr (15.05 cycle d$^{-1}$) based on the value reported by \citet{otulakowska2013ixdra}. The orbital period was estimated from long-term photometric observations using light curve analysis, O–C diagram, and power spectrum analysis. 

We detect a signal at 1.61 hr (14.91 cycle d$^{-1}$) along with its first harmonic in multiple sectors; however, given that the system exhibits frequent superoutbursts, the association of this signal with the orbital period remains uncertain.

\subsubsection*{\textbf{1RXS J064725.3+491539 (TIC 453331680)}}

The R\&K quotes a period of 1.58 hr (15.15 cycle d$^{-1}$) based on the value reported by \citet{littlefield2013vsnet15476_1RXS}. However, the original VSNET link is no longer accessible, and thus the method of determination is unknown.

We detected a strong signal at 1.75 hr (15.267 cycle d$^{-1}$) along with its first harmonic, both observed in two available sectors. We believe that this represents the true period of the binary.

\subsubsection*{\textbf{SDSS J110014.72+131552.0 (TIC 903453082)}}

The R\&K quotes 1.584 hr (15.15 cycle d$^{-1}$) based on the value reported by \citet{Kato2009vsnet11202}. Although the method used to determine this signal is unknown due to the inaccessibility of the original link, a superhump signal was reported for this source in \citet{kato2009superhumps}, corresponding to 1.62 hr (14.80 cycle d$^{-1}$). 

We detected a signal at 1.63 hr (14.735 cycle d$^{-1}$) in only one sector. But this signal occurred during an outburst. We don't think this signal is reliable to be the orbital period of the system.

\subsubsection*{\textbf{V*TV Crv (TIC 952240841)}}

The R\&K quotes 1.509 hr (15.91 cycle d$^{-1}$) based on the value reported in \citet{Woudt2003}. The orbital period was determined from high-speed photometric observations during quiescence, where the double-humped light curve produced alias structures in the Fourier transform that were resolved using the previously known superhump period as a constraint. 

We detected a strong signal at 1.518 hr. (15.814 cycle d$^{-1}$) and its first harmonic in two different sectors. We believe this to be the true orbital period of the binary.

\subsubsection*{\textbf{CRTS J134014.9-350512 (TIC 1055183341)}}

The R\&K quotes 1.416 hr (16.95 cycle d$^{-1}$) based on the value reported in \citet{Coppejans2014}. The orbital period was determined using either Fourier Transform (FT) or Phase Dispersion Minimization (PDM), depending on the shape of the system’s light curve. 

We detected a strong signal at 1.436 (16.72 cycle d$^{-1}$) and the first three harmonics. We believe this to be the true orbital period of the binary.

\subsubsection*{\textbf{V*IL Leo (TIC 840418301)}}
The R\&K quotes 1.367 hr (17.55 cycle d$^{-1}$) based on the value reported by \citet{Schmidt2007}. \citet{Schmidt2007} determined the orbital period by analysing the periodic modulations in the light curve using time-resolved photometry and the phase-folding method.

We detected a strong signal at 1.37 hr (17.4702 cycle d$^{-1}$). This signal is detected with high significance in two sectors. We believe this to be the true orbital period.

\subsubsection*{\textbf{TZ Per (TIC 347692846)}}

The R\&K catalogue quotes 6.31 hr (3.80 cycle d$^{-1}$) based on the radial velocity study of \citet{echevarra99}. Interestingly, \citet{ringwald95} also inferred a slightly different orbital period of 6.25 hr (3.84 cycle d$^{-1}$) through a radial velocity study. 
We recover a strong signal and related first harmonic at 6.00 hr (4.01 cycle d$^{-1}$). The same signal is present in both \tess\ sectors where this system was observed. Although TZ Per shows dwarf nova outbursts during the \tess\ observations, the 4.01 cycle d$^{-1}$, signal is present even when considering the quiescent data alone. It is unclear at this stage what the true orbital period for this system is, but we nonetheless retain this entry in our catalogue.





\subsubsection*{\textbf{2MASS J10395999-4701261 (TIC 146721954)}}

The R\&K quotes 3.785 hr (6.35 cycle \ensuremath{\mathrm{d}^{-1}}) based on the value reported by \citet{pretorius2008_hALPHA}. The orbital period was determined using radial velocity measurements derived from time-resolved spectroscopy of the H$\alpha$ emission line. 

We detected a strong signal at 3.77 hr ($6.37\ \mathrm{cycle}\ \mathrm{d}^{-1}$). We don't see any peak from \citet{pretorius2008_hALPHA} but our detected first fundamental frequency matches with the superhump period detected in \citet{Stefanov2023_TiltedDiscs}. 

\subsubsection*{\textbf{V* V751 Cyg (TIC 356777168)}}

The R\&K quotes 3.467 hr (6.92 cycle d$^{-1}$) based on the value reported by \citet{Patterson2001_V751Cygni}. The orbital period was determined using radial velocity measurements. 

We found a signal at 3.34 hr (7.19 cycle d$^{-1}$), but the same value is reported by \citet{Patterson2001_V751Cygni} as a negative superhump. We believe that the first fundamental here is a negative superhump. 

\subsubsection*{\textbf{V* LN UMa (TIC 103605968)}}

The R\&K quotes 3.47 hr (6.92 cycle d$^{-1}$) based on the value reported by \citet{Hillwig1998_PG1000}. The orbital period was determined from radial velocity measurements of the H$\beta$ emission line. This value is detected in TESS sector 47 observations. 
We detected a signal at 3.315 hr (7.246 cycle d$^{-1}$) but this value is reported as a negative superhump in \citet{Bruch2023}. We also accept this value as a negative superhump.

\subsubsection*{\textbf{V* V378 Peg (TIC 432280211)}}

The R\&K quotes 3.325 hr (7.215 cycle d$^{-1}$) based on the value reported by \citet{RINGWALD2012433}. The orbital period was determined  through a radial velocity study, derived from Doppler shift measurements of spectral lines. This study also reports a superhump period of 3.23 hr (7.41 cycle d$^{-1}$). 

\citet{Kozhevnikov2012V378Peg} interpret the 3.238\,h oscillation in V378~Peg as a permanent superhump as well and state that the orbital period could not be determined, and therefore the superhump cannot be categorized.

We found a strong signal at 3.243 hr (7.394 cycle d$^{-1}$). Within the stated uncertainties, we consider this value to be consistent with the negative superhump period reported in \citet{RINGWALD2012433} and \citet{Kozhevnikov2012V378Peg}. 

But an alternative explanation remains plausible.
The observed modulation could in fact be the orbital period, with the apparent phase and amplitude variability simply reflecting flickering that masks its coherence and produces superhump-like signatures in the power spectrum. This interpretation will require further study.




\subsubsection*{\textbf{CRTS J120052.9-152620 (TIC 180261612)}}	

The R\&K quotes 2.23 hr (10.75 cycle d$^{-1}$) based on the value reported by \citet{Masi2011}. In our analysis, this source was observed in only a single TESS sector, where our algorithm detected a strong signal at 2.14 hr (11.22 cycle d$^{-1}$). Due to the limited coverage from only one sector and the lack of sufficient information in the existing literature, we have decided to include this frequency in our summary table. However, we note that its classification as the true orbital period remains uncertain and should be treated with caution.

\subsubsection*{\textbf{CRTS J112815.4-344807 (TIC 942248803)}}	

The R\&K quotes an orbital period of 2.36 hr (10.15 cycle d$^{-1}$) based on  the value reported by \citet{Coppejans2014}. This value was derived from fitting a sinusoidal model to the light curve and identifying the dominant peak in the Fourier Transform (FT).

We found a strong signal at 10.44 cycle d$^{-1}$, (2.30 hr), along with a clearly visible first harmonic. The system was observed in TESS Sector 63 and Sector 90. Sector 63 shows clear evidence of an outburst, while Sector 90 captures the system in quiescence. The same dominant frequency is recovered in both sectors, supporting the interpretation that the 10.44 cycle d$^{-1}$, signal likely represents the orbital period. 

\subsubsection*{\textbf{V* CZ Aql (TIC 3399307)}}

The R\&K quotes an orbital period of 4.82 hr (4.98 cycle d$^{-1}$) based on the value reported by \citet{Sheets2007_CZ_Aqr} though they noted uncertainty due to aliasing, with plausible periods ranging between 4.798 and 4.826 hours. \citet{Bruch2017_CZ_Aqr} conducted time-resolved photometry of CZ Aql and reported a dominant periodicity at 5.2083 hours, likely affected by aliasing and strong flickering activity. 
We detect a coherent signal at 4.78 hr (5.029 cycle d$^{-1}$) in sector 80, consistent with the spectroscopic orbital period reported by \citet{Sheets2007_CZ_Aqr}, and a secondary signal near 4.62 hours (5.20 cycle d$^{-1}$) at sector 54. We believe the signal at 4.78 hr represents ther eal orbital period of the binary.

\subsubsection*{\textbf{V* ES Dra (TIC 202507376)}}

The R\&K quotes an orbital period of 4.238 hr (5.66 cycle d$^{-1}$) based on time-resolved photometry and phase-folding of the light curve. 

We detect a signal at 4.263 hr (5.634 cycle d$^{-1}$) and its related first 4 harmonics. We believe this value might be the true orbital period of the system.

\subsubsection*{\textbf{SDSS J1607.02+3623 (TIC 1200808784)}}

The R\&K quotes an orbital period of 3.49 hr (6.87 cycle d$^{-1}$) based on the value reported by \citet{ZenginCamurdan2010}. The orbital period was determined using time-resolved CCD photometry. 

We detect a strong signal at 3.75 hr (6.393 cycle d$^{-1}$) and its first 4 harmonics. Although only data from Sector 78 were used in this analysis, the presence of the first four harmonics supports the interpretation that this value might represent the true orbital period.

\subsubsection*{\textbf{V* KZ Gem (TIC 387612744)}}

The R\&K quotes an orbital period of 2.67 hr (8.99 cycle d$^{-1}$) based on the value reported by \citet{armstrong2015k2varcat_KZ_GEM}.However, we found a signal at 4.493 cycle d$^{-1}$, corresponding to 5.34 hr. We suspect the previous value is the first harmonic of the true orbital period of the binary, as also suggested by \citet{Dai_2020_KZ_gem}.

\subsubsection*{\textbf{HalphaJ130559 (TIC 253410027)}}

The R\&K quotes an orbital period of 3.928 hr (6.10 cycle d$^{-1}$) based on the value reported by \citet{pretorius2008_hALPHA}. The orbital period was determined through time-resolved spectroscopy by analysing radial velocity curves and identifying the dominant frequency in a Fourier transform. 

We detect a strong signal at 3.62 hr (6.626 cycle d$^{-1}$) and its first 6 harmonics in two \textit{TESS} sectors(64, 65). We also identify a positive superhump signal at 6.89 cycle d$^{-1}$, in both sectors. We suspect 3.62 hr signal is the true orbital period of the binary and the frequency reported in R\&K is likely affected by aliasing due to limited temporal coverage.

\subsubsection*{\textbf{2MASS J07491041-0549259 (TIC 426056469)}}

The R\&K quotes an orbital period of 3.6 hr (6.66 cycle d$^{-1}$) based on the value reported by \citet{Motch1998}. 2MASS J07491041-0549259 was identified as a polar cataclysmic variable by \citet{Motch1998} during a systematic optical follow-up of ROSAT Galactic Plane Survey sources. 

We detect a strong, coherent signal at 3.69 hr (6.519 cycle d$^{-1}$), along with its first three harmonics. We suspect this is the true orbital period of the binary. 




\subsubsection*{\textbf{MisV1448 (Gaia DR2 2234058025343096448) (TIC 233264291)}}

The R\&K quotes 4.329 hr (5.46 cycle d$^{-1}$) based on observations during the outburst of \citet{denisenko2012vsnet_MIS}. The period was determined using the Lafler-Kinman and Deeming methods. We find a period of 4.661 cycle d$^{-1}$, (5.15 hr) and the related first harmonic. We suspect the previous period determination is the positive superhump of the system, and we here quote the true orbital period.

\subsubsection*{\textbf{2MASS J19225496+4309059 (TIC 159448831)}}

The R\&K quotes 1.791 hr (13.40 cycle d$^{-1}$) based on the value reported by \citet{Kato2013KIC7524178}. The orbital period was determined by phase-dispersion minimization (PDM) analysis, while the superhump periods were found to range between 13.67–13.82 cycle d$^{-1}$ (negative) and 12.69–12.95 cycle d$^{-1}$ (positive).

We detected a signal at 1.747 hr (13.7518 cycle d$^{-1}$) which is consistent with the dominant negative superhump frequency (13.7–13.8 cycle d$^{-1}$) reported by \citet{Kato2013KIC7524178}. 

\subsubsection*{\textbf{RX J1600.2+7050 (TIC 1201303785)}}

The R\&K quotes 1.705 hr (14.08 cycle d$^{-1}$) based on the value reported by \citet{ASASSN13bj_VSNET_2013}. The other paper in R\&K \citep{kato2014pasj66_90} reported a period at 1.754 hr (13.68 cycle d$^{-1}$) by using the PDM method to determine the period. 

We detected a strong signal at 1.663 hr (14.43 cycle d$^{-1}$) and its first harmonic. We consider this signal to represent the true period of the binary. 

\subsubsection*{\textbf{J1735+1547 (TIC 267353881)}}
The R\&K quotes 83.95 min based on \citet{Kato_2013}. They used the Phase Dispersion Minimization (PDM) method to determine the orbital period. 
We find a period of 8.455 hr (5.6701 cycle d$^{-1}$) and the related first harmonic. We suspect this is the true orbital period of the binary. Further evidence for a long period CV in this system is also found by inspecting the Gaia CMD and finding this system close to the main sequence as several other long period CVs. We are unsure what causes the signals of \citet{Kato_2013}.

\subsubsection*{\textbf{2MASS J21543365+3550176 (Var 79 Peg) (TIC 399872907)}}

The R\&K quotes 1.53 hr (15.69 cycle d$^{-1}$) based on \citet{vanmunster2004_var79peg}. They used unfiltered time-series CCD photometry to detect light curve modulations. We find a period of 4.39 hr (5.464 cycle d$^{-1}$) and the related first harmonic. We suspect this is the true orbital period of the binary. 

\subsubsection*{\textbf{SDSS J134441.83+204408.3 (TIC 1000324193)}}

The R\&K quotes 1.67 hr (14.40 cycle d$^{-1}$) based on the value reported by \citet{szkody14}. The orbital period was determined by analysing radial velocity variations of the H$\alpha$ and H$\beta$ lines, supported by the presence of hump features in the photometric light curves. 

SDSS J1344+20 was initially considered to be a Polar, as \citet{szkody14} identified cyclotron hump features, large Balmer-line radial velocity amplitudes, and photometric hump structures consistent with accretion poles in magnetic cataclysmic variables. However, \textit{TESS} observations and ground-based facilities demonstrated that the system is in fact a highly asynchronous magnetic cataclysmic variable, with a spin-to-orbit ratio of $P_{\mathrm{spin}}/P_{\mathrm{orb}} \approx 0.893$, and a surface magnetic field strength of $56 \pm 2$ MG \citep{Littlefield2023SON}. \citet{Littlefield2023SON} reports that SDSS J1344+20 exhibits a white dwarf spin period of about 102 minutes (14.16 cycle d$^{-1}$) and an orbital period of about 114 minutes (12.64 cycle d$^{-1}$), based on TESS observations.

The Algorithm detected a strong signal at 1.695 hr (14.166 cycle d$^{-1}$). We consider this to be the spin period, as reported by \citet{Littlefield2023SON}, and take the orbital period to be 12.64 cycle d$^{-1}$.

\subsubsection*{\textbf{V* UV Gem (TIC 718010089)}}

The R\&K quotes 2.149 hr (11.17 cycle d$^{-1}$). However, we have not been able to identify the publication in which this entry was originally reported. \citet{kato2009superhumps} reported superhump periods in the range of 2.22–2.25 h (10.7–10.8 cycle d$^{-1}$) during the 2003 and 2008 superoutbursts. 
\citet{Dai2016_K2_CV_C0C1} report another orbital signal at approximately 11.32 cycle d$^{-1}$ (2.12 h) using a Lomb–Scargle analysis of the K2 C0 light curve.

We detected a strong signal at 2.20 hr (10.924 cycle d$^{-1}$),  which occured during an outburst. The algorithm also detected other signals at 10.94 cycle d$^{-1}$. The first detected signal falls within the expected superhump range in \citet{kato2009superhumps}. We believe that this signal corresponds to a positive superhump.

\subsubsection*{\textbf{BT CrB (TIC 1101672770)}}

The R\&K quotes 2.14 hr (11.21 cycle d$^{-1}$) based on the value reported by \citet{szkody2006cataclysmic}. The orbital period was determined from time-resolved spectra by measuring radial velocities with the double-Gaussian method and fitting a sinusoid to the velocity curve.

We detected a signal at 2.11 hr (11.36 cycle d$^{-1}$) and its first 3 harmonics. We observe this signal exclusively during the superoutburst and in a single sector; therefore, we do not confirm it as the orbital period, instead believing it to be a superhump signal.

\subsubsection*{\textbf{2MASS J16595166+1927454 (TIC 345765912)}}

The R\&K quotes 3.384 hr (7.09 cycle d$^{-1}$) based on the value reported by \citet{Thorstensen2015SDSS}. The orbital period in \citet{Thorstensen2015SDSS} determined by fitting sine curves to radial velocity measurements obtained from time-series spectroscopy. 

We find a strong period at 3.41 hr (7.044 cycle d$^{-1}$) and its first 3 harmonics at sector 25. \citet{Bruch2024_SYNOPTIC} refined the orbital period to 3.41 hr based on photometric data and reported, for the first time, that the system exhibits clear eclipses in the folded light curve. \citet{Bruch2024_SYNOPTIC}'s findings are fully consistent with our results and further support our interpretation that our period represents the orbital period.

\subsubsection*{\textbf{V* MV Lyr (TIC 158318859)}}

The R\&K quotes 3.1896 hr (7.52 cycle d$^{-1}$) based on the value reported by \citet{Skillman1995_MVLyr}. The orbital period was derived from H$\alpha$ emission-line radial velocity measurements. 

We detected a signal at  3.08 hr (7.7935 cycle d$^{-1}$). The superhump period range reported in \citet{bruch2022} confirms our identification of the detected signal as a negative superhump.




\subsubsection*{\textbf{RX J1643.7+3402 (TIC 224325028)}}

The R\&K quotes 2.895 hr (8.29 cycle d$^{-1}$) based on the value reported by \citet{Patterson2002_V442Oph_RXJ1643}. The orbital period was determined by time-resolved spectroscopy for measuring H$\beta$ emission-line radial velocity.

We detected a signal at 2.807 hr (8.55 cycle d$^{-1}$). But this signal was reported as a negative superhump signal in \citet{Patterson2002_V442Oph_RXJ1643}.



\subsubsection*{\textbf{V* TY PsA (TIC 47466200)}}



The R\&K quotes 2.02 hr (11.86 cycles d$^{-1}$) based on the value reported by \citet{Hamilton2011}. \citet{Hamilton2011} observed TY~PsA with medium-resolution K-band spectroscopy using VLT/ISAAC. No clear absorption features of the secondary were detected; however, continuum slopes at the red end indicated water vapor absorption, suggesting a late-type donor star (>M6).
Our algorithm detected a signal at 2.08 hr (11.56 cycles d$^{-1}$) based solely on the periodogram analysis of the light curve in sector 69. Although a signal at 11.86 cycles d$^{-1}$, consistent with the orbital period suggested by \citet{Hamilton2011}, is present but weak in the periodogram, its second harmonic at 23.74 cycles d$^{-1}$ appears much stronger. The 11.56 cycles d$^{-1}$ signal, which dominates the first half of the light curve during quiescence and vanishes in the second half, may therefore represent a transient positive superhump rather than the true orbital modulation.

\subsubsection*{\textbf{V* WY Tri (TIC 620949163)}}

The R\&K quotes 1.82 hr. (13.17 cycle d$^{-1}$) based on the value reported by \citet{Vanmunster2001}. However, since the corresponding link is no longer accessible, we do not know by which method it was determined. Another reference related to this source is \citet{kato2009superhumps} but there is not any orbital period reported. 

We detected a signal at 1.81 hr (13.237 cycle d$^{-1}$) and its first harmonic. The values reported in the literature also support this result within the margin of error \citep{WY_Tri1, WY_Tri2}. We consider this to be the true period of the system.



\subsubsection*{\textbf{DW Cnc (TIC 19028616)}}

The R\&K quotes 1.43 hr (16.73 cycle d$^{-1}$) as spectroscopic orbital period \citep{Patterson2004PASP..116..516P,Rodriguez-Gil2004MNRAS.349..367R}. We find multiple coherent signals corresponding to the system's spin (37.32 cycle d$^{-1}$) and a negative superhump (16.9299 cycle d$^{-1}$) along with their beats and harmonics. These are listed and explained in detail in Table 3 in \citet{Veresvarska2025MNRAS.539.2424V}. The orbital period is not detected here, most likely due to the low inclination of the system. 

\subsubsection*{\textbf{V* V1006 Cyg (TIC 352302825)}}

The R\&K quotes 2.38 hr (10.08 cycle d$^{-1}$) based on the value reported by \citet{V1006_CYG}. The orbital period was determined by fitting radial-velocity measurements from different nights and resolving the cycle count to find the best-fitting sinusoidal solution. 

We detected a signal near 20.195 cycle d$^{-1}$ in four different sectors, but we believe this signal is the harmonic of the signal reported by \citet{V1006_CYG}. 

\subsubsection*{\textbf{DM Gem (TIC 764302436)}}

The R\&K catalogue lists two periodicities for DM~Gem, namely 
0.12287~d and 0.12266~d (2.95~hr and 2.94~hr), corresponding to 
8.14 and 8.15~cycle~d$^{-1}$, respectively, based on the Wise Observatory 
CCD photometry reported by \citet{Lipkin_DMGem_Report}. These signals were 
interpreted as closely spaced quasi-periodic oscillations rather than a 
stable orbital modulation.

We find a signal at  8.641 cycle d$^{-1}$ detected in three different sectors, with its first  harmonic also visible in Sector 72. However, given its proximity to the  QPO frequencies reported by \citet{Lipkin_DMGem_Report} and the lack of  coherent behaviour across sectors, we do not consider the  8.641 cycle d$^{-1}$ signal to be a secure orbital period candidate for DM Gem.

\subsection*{Systems below the period minimum}

\subsubsection*{\textbf{EI Psc (TIC 423324616)}}

EI Psc is one of the well-known hydrogen-rich CVs below the period minimum with an orbital period of 0.044566904(6) d \citep{Thorstensen_2017, Thorstensen2002}.
We found a period of 22.439 cycles d$^{-1}$ in the TESS observations in Sectors 42, 56, 70, and 83, which is consistent with the known orbital period. No superoutbursts were observed during these TESS observations.

\subsubsection*{\textbf{Gaia DR3 875941782403757056 = SDSS J075233.17$+$294339.7 (TIC 740684485)}}

SDSS J075233.17$+$294339.7 was suggested to be a CV candidate in VSNET-chat 7967\footnote{\url{http://ooruri.kusastro.kyoto-u.ac.jp/mailarchive/vsnet-chat/7967}}. 
The TESS light curve shows short ($\leq$1 d) bursts with a $\simeq$10\% amplitude. The light curve shows a periodicity at 19.967 cycle d$^{-1}$, consistent between the two sectors 60 and 72. The folded light curve with the half frequency still shows equal minima and maxima, suggesting that the 19.967 cycle d$^{-1}$ is likely a true period.

\subsubsection*{\textbf{MGAB-V249 = UCAC4 672-097665 = ZTF J2130+4420 (TIC 240326669)}}

MGAB-V249 was first identified as an eclipsing AM~CVn candidate with an orbital period of 39.3401(1) min, corresponding to a frequency of 36.60 cycle d$^{-1}$, based on ZTF observations \citep{Kupfer_2020}. Subsequent spectroscopic follow-up demonstrated that the system is not a classical AM~CVn binary, but instead consists of a Roche-lobe-filling helium-rich hot subdwarf donor transferring mass to a white dwarf \citep{Kupfer_2020}.Our analysis of the \textit{TESS} light curves recovers a dominant frequency at 36.6 cycle d$^{-1}$, corresponding to an orbital period of approximately 39.3 min, in agreement with the literature value.

\subsubsection*{\textbf{Gaia14aae (ASASSN-14cn) (TIC 1201247611)}}

Gaia14aae was first identified following its outburst in 2014 and was subsequently confirmed as an eclipsing AM~CVn system with an orbital period of 49.71 min \citep{Camplbell}.
Part of the \textit{TESS} photometric data for this source has previously been presented by \citet{PichardoMarcano2021}. Our analysis of the \textit{TESS} light curves reveals a consistent periodicity at 28.969 cycle d$^{-1}$, fully consistent with the known orbital period of the system.

\subsubsection*{\textbf{DDE 170 (TIC 29290702)}}

DDE 170 was identified as a variable star by CRTS, and is also registered in the SDSS WD-MS catalog from \citet{Rebessa}.
The system shows low and high states in ZTF\footnote{\url{https://alerce.online/object/ZTF18aagrcdr}}, suggesting a Polar classification. 
The frequency 26.6 cycle d$^{-1}$ is present in all available TESS sectors. The phase-folded profile with the half frequency shows uneven minima, which suggests that the true orbital frequency can be half of the detected one; 13.3 cycle d$^{-1}$. 



\subsubsection*{\textbf{Gaia DR3 344970334401056512 = CRTS J020254.8$+$403703 (TIC 621589134)}}


CRTS J020254.8$+$403703 was first identified as a dwarf nova candidate in \citet{Coppejans}.
Even the outburst maximum is below 18 mag, which is around the typical limiting magnitude of TESS. Since there are several brighter systems within 30", the detected period of 19.63 cycle d$^{-1}$ can be unrelated to this system or contaminated by nearby systems.

\subsection*{Magnetic CVs}

\subsubsection*{\textbf{Gaia 19ayl (TIC 64107454)}}
Gaia 19ayl was discovered as a CV candidate by \textit{Gaia}. It is listed in the AAVSO VSX database as a polar candidate, with a reported magnitude range of 16.8-20.8 in the AAVSO CV band.  No information regarding its orbital period is currently available in the literature. The object was observed in only one \textit{TESS} sector, where our automated analysis detected a strong signal at 13.298 cycle d$^{-1}$ (1.805 hr), which may correspond to the orbital period of the system.

\subsubsection*{\textbf{ASASSN-16pm (TIC 660149869)}}
ASASSN-16pm is listed in the AAVSO VSX database as a polar candidate, with a reported magnitude range of 14.1-21.0 in the AAVSO-CV band.  The database also lists a frequency of 6.496 cycle d$^{-1}$, derived from eclipse minima timings, although we could not find a peer-reviewed publication confirming this value. ASASSN-16pm  was observed in only one \textit{TESS} sector, where this signal is clearly detected in the data.

\subsubsection*{\textbf{1RXS J061607.6$+$745217 (TIC 705223072)}}
1RXS J061607.6$+$745217 is listed in the AAVSO VSX database as a polar candidate, with a reported magnitude range of 18.3-20.6 in the ZTF-g band.  The database also lists a frequency of 10.68 cycle d$^{-1}$, although we could not find a peer-reviewed publication confirming this value. 1RXS J061607.6$+$745217 was observed in two \textit{TESS} sectors, where the signal at 10.687 cycle d$^{-1}$ is clearly present in both sectors. Since this frequency was recovered independently in each sector, we adopt it as the first fundamental frequency in our catalogue.

\subsubsection*{\textbf{ATO J110.1217+49.2737 (TIC 150028639)}}
ATO J110.1217+49.2737, also known as Khrapov 2, is listed in the AAVSO VSX database as a polar candidate, with a reported magnitude range of 16.3-19.0 in the AAVSO-CV band. No information regarding its orbital period is currently available in the literature. The object was observed in two \textit{TESS} sectors, where our automated analysis detected a strong signal at 6.603 cycle d$^{-1}$, which we adopt as the first fundamental frequency in our catalogue.

\subsubsection*{\textbf{ZTF18abujfcu (TIC 743283324)}}
ZTF18abujfcu is listed in the AAVSO VSX database as a polar candidate based on its long bright state in the ZTF data reported by \citep{Kato2022}. 
An optical spectrum obtained by \citet{2024AJ....167..186S} showed features consistent with a polar classification. No information regarding its orbital period is currently available in the literature. The object was observed in two \textit{TESS} sectors, where our automated analysis detected a strong signal at 14.821 cycle d$^{-1}$, which we adopt as the orbital period in our catalogue.

\subsubsection*{\textbf{DDE 37 (TIC 2020916168)}}
DDE 37 was discovered as a variable star by \citep{Denisenko2025} and is listed in the AAVSO VSX database as a polar candidate based on its high and low states.  No information regarding its orbital period is currently available in the literature. The object was observed in two \textit{TESS} sectors, where our automated analysis detected a strong signal at 13.663 cycle d$^{-1}$, which we adopt as the first fundamental frequency in our catalogue. Visual inspection revealed two additional low-amplitude peaks at 40.98 cycle d$^{-1}$ and 54.64 cycle d$^{-1}$, corresponding to the third- and fourth-order harmonics of the fundamental frequency.


\subsubsection*{\textbf{1RXS J232928.0$-$161654 (TIC 328010691)}}
1RXS J232928.0$-$161654 was discovered as a transient by ZTF and identified as a likely polar by \citep{kato2021} based on its long bright states with high-amplitude variations in ZTF data.  No information regarding its orbital period is currently available in the literature. The object was observed in two \textit{TESS} sectors, where our automated analysis detected a strong signal at 19.023 cycle d$^{-1}$, which we adopt as the first fundamental frequency in our catalogue. Visual inspection also revealed three additional peaks at 18.56 cycle d$^{-1}$, 18.80 cycle d$^{-1}$, and 19.24 cycle d$^{-1}$.

\subsection*{Notes on Individual Systems}

\subsubsection*{\textbf{V* DT Oct (TIC 277074217)}}

In sector~39 we detected a signal at 13.355 cycle d$^{-1}$ and its harmonic that emerge simultaneously with the superoutburst.  
The same signal was also identified in sectors~27. We do not consider this signal to represent the orbital period of the system.  
As reported by \citet{LiuQian2024_TESS_SUUMa}, TESS data place the superhump frequencies of the system in the range of 13.01–13.37 cycle d$^{-1}$, with a period of 13.75 cycle d$^{-1}$.  
Additional peaks returned by our algorithm at 13.18 and 13.39 cycle d$^{-1}$, fall within this superhump range, and we therefore interpret the 13.35 cycle d$^{-1}$ signal as a superhump.

\subsubsection*{\textbf{CRTS CSS140309 J102844-161303 (TIC 334693154)}}

CRTS CSS140309 J102844-161303 was observed in three \textit{TESS} sectors, and our algorithm consistently detected a strong signal at 
10.517 cycle d$^{-1}$, in all three sectors. Since this frequency was recovered independently in each sector and the first fundamental signal algorithm detected, it is adopted as the first fundamental frequency in our catalogue. However, in Sector~35 we also detect a weaker peak at 5.20 cycle d$^{-1}$, raising the possibility that this signal is the true fundamental and that the 10.517 cycle d$^{-1}$ peak corresponds to its first harmonic.

\subsubsection*{\textbf{V* V373 Cen (TIC 72182461)}}

V373~Cen was observed in two \textit{TESS} sectors.
The orbital frequency of V373 Cen is listed in the AAVSO VSX database as 2.79 cycle d$^{-1}$, but we could not identify a peer-reviewed publication reporting this value. This signal is not detected in the \textit{TESS} data.
Instead, we identify a strong peak at 3.7535 cycle d$^{-1}$ in both sectors, which appears with outburst.
When the pre–outburst quiescent data of Sector~37 are analysed separately, a very weak signal at 1.89 cycle d$^{-1}$ emerges.
This lower frequency may represent the true orbital period, with the 3.7535 cycle d$^{-1}$ peak corresponding to its first harmonic. 

\subsubsection*{\textbf{V* V822 Cen (TIC 461646008)}}

V822~Cen was observed in two \textit{TESS} sectors.
Our automated analysis consistently detected a strong signal at 3.179 cycle d$^{-1}$, in both sectors. However visual inspection of the light curves reveals a weaker peak near 1.59 cycle d$^{-1}$ in each sector. Because this lower-frequency signal was not recovered by the detection algorithm it is not included in the catalogue, but it may represent the true fundamental frequency, with the 3.179 cycle d$^{-1}$ peak corresponding to its first harmonic. This frequency is consistent with values reported in the literature \citet{Cowley1988CenX4}.

\subsubsection*{\textbf{ASASSN -20fg (TIC 395719608)}}

ASASSN-20fg was observed in two \textit{TESS} sectors.
Our automated analysis consistently detected a strong signal at 8.397 cycle d$^{-1}$, in both sectors. But visual inspection of Sector~83 reveals an additional peak near 4.20 cycle d$^{-1}$. This lower–frequency signal is not recovered by the detection algorithm, but it may represent the true fundamental frequency, with the 8.397 cycle d$^{-1}$ peak corresponding to its first harmonic.

\subsubsection*{\textbf{V* V1147 Cen (TIC 404153438)}}

V1147~Cen was observed in two \textit{TESS} sectors, both of which display outburst activity. Our algorithm detected a strong signal at 4.766 cycle d$^{-1}$, in both sectors and recorded it as the first fundamental frequency. However, when the pre–outburst quiescent data of Sector~64 are analysed separately, a strong signal at 2.38 cycle d$^{-1}$ becomes evident. We consider this frequency to be the true first fundamental, with the 4.766 cycle d$^{-1}$ peak corresponding to its first harmonic.





\subsubsection*{\textbf{Gaia DR3 2238460783433624832 = MGAB-V248 (TIC 1883998117)}}

MGAB-V248 was first identified as a variable star with a period of 27.95 min (51.51 cycle d$^{-1}$) in ZTF\footnote{\url{https://sites.google.com/view/mgab-astronomy/mgab-v201-v250-hidden}}. \citet{Gavin} suggested that this period can be attributed to the rotation of the DAB white dwarf rather than an AM CVn star, since its spectrum shows hydrogen and helium absorption lines. 
We detected a periodicity at 0.466 hr (51.51 cycle d$^{-1}$), consistent with the above-mentioned period, while more observations (i.e., radial velocities) are needed to confirm if this period is orbital or not.

\subsubsection*{\textbf{ZTF17aaburxr (TIC 375982881)}}

ZTF17aaburxr was observed by \textit{TESS} in two sectors. The algorithm detected two apparent fundamental frequencies at 2.048~cycle~d$^{-1}$ and 14.51~cycle~d$^{-1}$ in Sector~77, while no significant signal is present at either frequency in Sector~78. In Sector~77, the first harmonic of the 14.51~cycle~d$^{-1}$ signal is visible near 29.04~cycle~d$^{-1}$. However, the 2.048~cycle~d$^{-1}$ peak is most likely caused by blending with a nearby bright W~UMa-type eclipsing binary, GSC~04476$-$00910 \citep{VSX_621217}, located about 2~arcmin from the CV. This interloping source has an orbital period of 0.976084~d ($\Omega = 1.03$~cycle~d$^{-1}$), producing a strong signal at its second harmonic (2$\Omega \approx 2.06$~cycle~d$^{-1}$), consistent with the detected frequency. Therefore, we interpret the 2.048~cycle~d$^{-1}$ signal as contamination, while the 14.51~cycle~d$^{-1}$ signal and its harmonic likely originate from the CV itself.

\subsubsection*{\textbf{OGLE MC-DN-30 (Gaia16all) (TIC 735229757)}}

\citet{PichardoMarcano2021} analysed Gaia 16all using 2-minute cadence \textit{TESS} light curves, detrending the superoutburst plateau and applying both Lomb--Scargle and Phase Dispersion Minimization (PDM) techniques. With this method, they obtained a candidate orbital period of $30.14 \pm 0.07$~min ($\approx$47.7~cycle~d$^{-1}$), consistent with the value predicted from the superoutburst recurrence time ($\sim$32~min) using the empirical relation of \citet{Levitan2015MNRAS}.

Our algorithm detected a signal near 3.093~cycle~d$^{-1}$ and its harmonics. However, this frequency is almost certainly due to contamination from a nearby, brighter W~UMa-type eclipsing binary, ASASSN-V~J062722.08$-$751406.4 \citet{VSX_839280}, located only 0.56~arcmin from the CV. This interloping source has an orbital period of 0.323335~d ($\Omega = 3.093$~cycle~d$^{-1}$), producing a strong signal at both the fundamental and 2$\Omega$ harmonic. Therefore, we attribute the detected 3.093~cycle~d$^{-1}$ signal to blending with this eclipsing binary rather than intrinsic variability from the system itself. Therefore, this system was excluded from the final catalogue.

\subsubsection*{\textbf{V* V1432 Aql (TIC 242812889)}}

V1432 Aql is listed in \citet{Schwope2025PolarCat} as the only known asynchronous polar with a white-dwarf spin period longer than its orbital period, with a degree of asynchronism of 1.0028. The orbital period is reported as 0.1402349 d (7.13 cycle d$^{-1}$) and the white-dwarf spin period as 0.1406256 d (7.11 cycle d$^{-1}$). Because the difference between these two periods is extremely small, the TESS light-curve power spectra cannot resolve the orbital and spin signals as separate peaks \citep{Bruch2025_TESS_IPs}. In the two TESS sectors analysed here we detect a signal at approximately 7.1226 cycle d$^{-1}$. Since the literature values for the spin and orbital frequencies are about 7.11 cycle d$^{-1}$ and 7.13 cycle d$^{-1}$, respectively, we cannot unambiguously determine which of the two modulations this detection corresponds to and therefore report only the measured frequency.

\subsubsection*{\textbf{V* TW Vul (TIC 290769912)}}

In \citet{WY_Tri1}, the orbital period of TW~Vul is reported as 6.29~cycle~d$^{-1}$, determined from radial-velocity measurements. 
In our analysis of the \textit{TESS} light curve, we detect two significant signals at 4.864~cycle~d$^{-1}$ and 12.49~cycle~d$^{-1}$. 
The 4.864~cycle~d$^{-1}$ signal is, however, very likely caused by blending with a nearby bright (V~$\approx$~10) W~UMa-type eclipsing binary, 
\textit{ASAS~J203925+2715.2} \citep{ASAS_J203925+2715.2_VSX}, located about 2~arcmin from TW~Vul. 
This contaminating system has an orbital period of 0.410687~d (2.435~cycle~d$^{-1}$), producing a strong second harmonic at 4.870~cycle~d$^{-1}$, nearly identical to the detected 4.864~cycle~d$^{-1}$ signal in our data. 
The other detected signal at 12.49~cycle~d$^{-1}$ is not harmonically related to the W~UMa binary’s frequency and is therefore likely intrinsic to TW~Vul. 
The literature frequency at 6.29~cycle~d$^{-1}$ is not detected in the \textit{TESS} periodogram.

\subsubsection*{\textbf{Gaia 19bdy (TIC 464626077)}}

Gaia19bdy was observed in three TESS sectors. Our algorithm identified a first fundamental at 0.91 cycle d$^{-1}$ and a second fundamental near 7.237 cycle d$^{-1}$. The nearby source Gaia DR3 5255409772517628160 \citep{GaiaDR3_5255409772517628160_2025}, located 0.44 arcmin away, is an eclipsing binary with a period of 2.19927 d. Its 2$\Omega$ harmonic (0.909 cycle d$^{-1}$) coincides with the 0.91 cycle d$^{-1}$ signal detected in TESS, indicating that this low-frequency modulation most likely originates from contamination by the eclipsing binary. The higher-frequency signal at 7.237 cycle d$^{-1}$, however, cannot be explained by this source and is instead proposed to represent the orbital modulation of the CV itself. We note that the VSX lists seven variable stars within 1.3 arcmin of the target, suggesting significant blending in the TESS aperture, and therefore interpret our results with caution.

\subsubsection*{\textbf{V* GW Lib (TIC 225798235)}}

Spectroscopic studies of GW Lib (\citealt{Toloza2016GWLIB}; \citealt{Pala2021gwlib})
consistently place the orbital period at
\( P_{\mathrm{orb}} \simeq 0.053~\mathrm{d} \)
(\(\approx 76~\mathrm{min} \approx 18.9~\mathrm{c\,d^{-1}}\)).

In our \textit{TESS} Sector~38 analysis, we recover a first–fundamental
signal at
\( f_{1} = 11.73~\mathrm{c\,d^{-1}} \)
(\(\approx 2.04~\mathrm{h}\)).
When the Sector~38 light curve is divided into halves, a slightly different
peak near \(12~\mathrm{c\,d^{-1}}\) appears, and the full–sector periodogram
shows beat features between these two close frequencies.
Although this photometric fundamental lies well below the spectroscopic
orbital frequency, \citet{Pala2021gwlib} and \citet{Hakala2025GWLIB}
demonstrate that GW\,Lib–type systems can experience short term,
accretion driven brightenings and mode changes, processes capable of
producing temporary frequency shifts or apparent fundamentals in the
\textit{TESS} bandpass. Furthermore, this signal is interpreted in \citet{veresvarska24} as a Quasi-periodic Oscillation (QPO) caused by a magnetically warped, precessing accretion disk. 



\subsubsection*{\textbf{V* FL Cet (TIC 422651186)}}

In our TESS analysis we detect a strong first fundamental at 33.04 $\mathrm{c\,d}^{-1}$ (43.6 min). However, the orbital period of 87 min (16.55 $\mathrm{c\,d}^{-1}$) is well established in the literature \citep{Schwope2025PolarCat, Hernandez_2025, Mason2015}.
The discrepancy arises because of strong cyclotron beaming, which produces two brightness maxima per true orbital cycle.

\subsubsection*{\textbf{WW Hor (TIC 166803096)}}

The algorithm detected a strong signal at 24.8 $\mathrm{c\,d}^{-1}$ (58 min).
However, the orbital period of 115 min (12.52 c/d) is firmly established in the literature \citep{Schwope2025PolarCat, Hernandez_2025, Ramsay2001_WWHor}. The algorithm detected the first harmonic. 

\subsubsection*{\textbf{1RXS J150618.6-750157 (TIC 403018318)}}

The source 1RXS~J150618.6$-$750157 was observed by \textit{TESS} in two sectors. 
In both sectors the algorithm detected a strong signal near $12.75~\mathrm{c\,d^{-1}}$,  
whereas visual inspection revealed an additional but much weaker feature around $6.37~\mathrm{c\,d^{-1}}$  
that was not identified by the algorithm owing to the signal–to–noise ratio in its vicinity.  
We therefore consider the $6.37~\mathrm{c\,d^{-1}}$ signal to represent the orbital period of the system.

\subsubsection*{\textbf{V* V523 Lyr (TIC 1876766376)}}

\citet{Mason2016} reported an orbital frequency of $6.32~\mathrm{c\,d^{-1}}$ from \textit{Kepler} photometry, 
which was constrained and confirmed through radial velocity spectroscopy obtained with the Hale 5.1\,m telescope. 
They also identified a superhump signal at 6.62~c/d. 

Our algorithm detects a strong signal at $7.167~\mathrm{c\,d^{-1}}$ in the \textit{TESS} data, 
while neither the reported orbital frequency nor the superhump frequency is clearly present in these observations.

\subsubsection*{\textbf{ASASSN -15bi (TIC 186289247)}}

ASASSN-15bi was observed by TESS in two sectors (Sectors 91 and 65). In both sectors, the algorithm detected a signal at 8.868~cycles~day$^{-1}$. However, once the outburst data are removed from the Sector 65 light curve, a signal at approximately 4.33~cycles~day$^{-1}$ becomes clearly visible. We interpret this lower-frequency signal as the orbital modulation, with the 8.868~cycles~day$^{-1}$ peak detected by the algorithm corresponding to its second harmonic.

\subsubsection*{\textbf{2MASS J22063080-5244193 (TIC 143853116)}}

2MASS J22063080--5244193 is observed by TESS in three sectors (95, 68, and 28), and an outburst is present in all of them. The algorithm detects a weak signal at 12.279~cycle~d$^{-1}$ in Sectors~28 and~68. However, when using only the quiescent data in these sectors a faint signal appears near 6.12~cycle~d$^{-1}$. We consider this to be the first fundamental frequency, with the algorithm's detected value likely corresponding to its second harmonic.

\subsubsection*{\textbf{ASASSN -14kv (TIC 128070524)}}

ASASSN-14kv is observed by TESS in two sectors (73 and 60). The algorithm detects a strong signal at 5.111~cycle~d$^{-1}$ in both sectors. However, during visual inspection, an additional weak feature appears near 2.55~cycle~d$^{-1}$, very close to the noise level. We consider this weaker signal to be the first fundamental frequency, with the detected 5.111~cycle~d$^{-1}$ signal likely representing its second harmonic.

\subsubsection*{\textbf{V* V478 Her (TIC 362131340)}}

V*~V478~Her is observed by TESS in five sectors. The algorithm detects a strong signal near 3.18~cycle~d$^{-1}$ in all sectors and records this as the first fundamental frequency. However, during visual inspection, we notice another feature of comparable strength around 1.57~cycle~d$^{-1}$. Because the noise level within approximately $\pm 0.3$~cycle~d$^{-1}$ of this frequency is relatively high, the algorithm does not identify this signal. We consider the 1.57~cycle~d$^{-1}$ feature to be the orbital frequency of the system, with the detected 3.18~cycle~d$^{-1}$ peak likely corresponding to its second harmonic.

\subsection*{\textbf{2MASS J05243042+4244506 (Paloma) (TIC 369210348)}}

Paloma (TIC 369210348) is an asynchronous magnetic cataclysmic variable exhibiting multiple coherent photometric periodicities associated with the white dwarf spin, the binary orbital motion, and their sideband (beat) frequencies. Such systems are known to display complex frequency structures in optical light curves, particularly when accretion occurs through magnetically channelled flows. The source is observed by TESS in only one sector (Sector 19).

\citet{Hernandez_2025} identify an orbital frequency of 9.1608 cycle~d$^{-1}$, corresponding to an orbital period of approximately 157 minutes \citep{Schwarz2007, Littlefield2023SON}. Our analysis detects a strong signal at 1.406 cycle~d$^{-1}$, which is recorded as the first fundamental frequency. However, this frequency is interpreted as the beat frequency of the system rather than the orbital modulation. Although the detected frequency does not correspond to the orbital period, we retain the 1.406 cycle~d$^{-1}$ signal as the first fundamental frequency in our catalogue, as it represents the dominant photometric modulation present in the TESS light curves.

\bsp	
\label{lastpage}
\end{document}